\documentclass[aps,prd,onecolumn,nofootinbib,superscriptaddress,longbibliography]{revtex4-2}

\usepackage{amsmath,amssymb}
\usepackage{mathrsfs}
\usepackage{xcolor}
\usepackage[colorlinks=true,linkcolor=blue,citecolor=blue,urlcolor=blue]{hyperref}
\usepackage{orcidlink}

\newcommand{\im}{\operatorname{Im}}
\newcommand{\re}{\operatorname{Re}}
\newcommand{\Qc}{\mathcal{Q}}
\newcommand{\GB}{\mathcal{G}}
\newcommand{\Po}{\mathcal{P}}
\newcommand{\Ss}{\mathcal{S}}
\newcommand{\Cc}{\mathcal{C}}

\begin{document}

\title{Gravitational memory, spin memory and S-duality at null infinity of the heterotic string}

\author{Maickol Mu\~noz-Palma\orcidlink{0009-0008-1017-4406}}
\email{m.munoz58@ufromail.cl}
\affiliation{Departamento de Ciencias F\'isicas, Universidad de La Frontera, Casilla 54-D, 4811186 Temuco, Chile}

\author{Alvaro Restuccia}
\email{alvaro.restuccia@uantof.cl}
\affiliation{Departamento de F\'isica, Facultad de Ciencias B\'asicas, Universidad de Antofagasta, Casilla 170, Antofagasta, Chile}

\author{Francisco Tello-Ortiz
\orcidlink{0000-0002-7104-5746}}
\email{francisco.tello@ufrontera.cl}
\affiliation{Departamento de Ciencias F\'isicas, Universidad de La Frontera, Casilla 54-D, 4811186 Temuco, Chile}

\date{\today}

\begin{abstract}
We determine the asymptotic structure at future null infinity of heterotic string theory compactified on $T^6$, in the consistent truncation to the Einstein-frame metric, the axion--dilaton $\tau=a+ie^{-2\phi}$ and the gauge fields descending from ten dimensions, through first order in $\alpha'$. Three results stand out. First, the radiative data, balance laws and memory observables organize under the S-duality group of the four-dimensional theory: the scalar flux in the Bondi mass loss is the invariant metric of $SL(2,\mathbb R)/U(1)$ evaluated on the scalar news, the gauge flux is the norm of a self-dual radiation field, and at fixed vacuum the residual duality rotation acts with charges $0$, $1$ and $2$ on gravitational, electromagnetic and dilaton--axion radiation, so that S-dual processes share their gravitational waveform and memories and differ in their scalar memory. Second, the spin memory receives a gauge-field contribution whose ordinary part is fixed by the Dirac--Schwinger--Zwanziger product of the total charge with the change of the charge aspect; it is generated by mutually nonlocal dyons, and its dipole part agrees in magnitude and sign with the angular momentum radiated by a charge scattered off a magnetic monopole. Third, within the four-derivative effective action, the Gauss--Bonnet and Pontryagin couplings fixed by the Green--Schwarz mechanism leave the mass-loss and angular-momentum balance laws unchanged at first order in $\alpha'$ and first enter the evolution of the spin-two curvature coefficient, $\partial_u\Psi_0^0\supset\tfrac{3i\alpha'}{2}\tau_1\bar\Psi_4^0$, which combines dilaton and axion into the complex scalar charge aspect and is not invariant under the residual duality rotation. For black holes with fixed physical charges $\tau_1$ is of order one, so this is a first-order effect in $\alpha'/M^2$.
\end{abstract}

\maketitle

%=====================================================================
\section{Introduction}
\label{sec:intro}

The asymptotic structure of isolated gravitating systems at null infinity, formulated by Bondi, van der Burg, Metzner and Sachs~\cite{Bondi:1962px,Sachs:1962wk}, revealed that the asymptotic symmetry group of asymptotically flat spacetimes is the infinite-dimensional Bondi--Metzner--Sachs (BMS) group. Its supertranslations are tied to the gravitational displacement memory~\cite{Zeldovich:1974gvh,Braginsky:1987kwo,Christodoulou:1991cr,Thorne:1992sdb} and to Weinberg's soft graviton theorem, the three vertices of the infrared triangle~\cite{Strominger:2013jfa,Strominger:2014pwa,Strominger:2017zoo}. Its extension by superrotations~\cite{Barnich:2009se,Kapec:2014opa,Campiglia:2014yka} is related in the same way to the spin memory, the persistent relative delay between counter-orbiting light rays produced by radiation that carries angular momentum~\cite{Pasterski:2015tva,Flanagan:2015pxa,Nichols:2017rqr,Compere:2018ylh,Mitman:2020pbt}. Memories are therefore not only prospective observables of gravitational-wave astronomy, but also direct probes of the infrared structure of the theory that governs the radiation.

This viewpoint has motivated the extension of memory effects and flux-balance laws to gauge fields~\cite{Bieri:2013hqa,Pasterski:2015zua,Pate:2017vwa}, to scalar fields and their dual two-forms~\cite{Campiglia:2018see,Francia:2018jtb,Seraj:2021rxd,Satishchandran:2019pyc}, and to theories beyond general relativity~\cite{Hou:2020tnd,Tahura:2020vsa,Hou:2021oxe,Hou:2021bxz,Heisenberg:2023prj}. The asymptotic solution space and the mass loss of Einstein--Maxwell--dilaton theory were obtained in Refs.~\cite{Lu:2019jus,Tan:2020hfa}, memory effects in theories with scalar couplings to the Gauss--Bonnet and Pontryagin densities were analyzed in Refs.~\cite{Heisenberg:2023prj,Hou:2021oxe,Hou:2021bxz}, and general conditions under which a theory with additional fields admits a smooth null infinity with BMS symmetry, together with the associated balance laws, were derived in Refs.~\cite{Maibach:2026afb,Maibach:2026jbf}. In most of these analyses the field content and the couplings are chosen as models. String theory instead fixes both, and endows the resulting low-energy theory with symmetries that have no counterpart in a generic model. Its massless sector is the natural arena in which to ask how an ultraviolet-complete theory of gravity organizes the infrared structure at null infinity~\cite{Avery:2015gxa,DiVecchia:2015oba}.

Heterotic string theory~\cite{Gross:1984dd,Gross:1985rr} provides a particularly sharp version of this question. Its toroidal compactification to four dimensions yields $\mathcal N=4$ supergravity coupled to vector multiplets~\cite{Maharana:1992my,Sen:1994fa}, whose bosonic sector contains the metric, the dilaton, the universal axion dual to the Kalb--Ramond field, the moduli and the gauge fields. The dilaton and the axion combine into a complex field $\tau=a+ie^{-2\phi}$ on $SL(2,\mathbb R)/U(1)$, and the two-derivative equations of motion are invariant under an $SL(2,\mathbb R)$ that acts on $\tau$ by M\"obius transformations and on the abelian gauge fields as an electric--magnetic duality rotation~\cite{Font:1990gx,Schwarz:1992tn,Ortin:1992ur,Sen:1994fa,Gibbons:1995ap}. This is the classical counterpart of S-duality. It is a symmetry of the four-dimensional theory itself, it relates electrically and magnetically charged black holes~\cite{Gibbons:1987ps,Garfinkle:1990qj,Shapere:1991ta,Kallosh:1992ii,Sen:1992ua}, and its spectrum contains mutually nonlocal dyons whose charges obey the Dirac--Schwinger--Zwanziger quantization condition~\cite{Dirac:1931kp,Schwinger:1966nj,Zwanziger:1968rs}. At the next order in the inverse string tension, the Green--Schwarz mechanism~\cite{Green:1984sg} fixes the couplings of the dilaton to the Gauss--Bonnet density and of the axion to the Pontryagin density~\cite{Zwiebach:1985uq,Gross:1986mw,Metsaev:1987zx,Bergshoeff:1989de,Campbell:1990fu,Kanti:1995vq,Cano:2019ycn}, which are the leading genuinely stringy corrections to the low-energy dynamics. The asymptotic analysis connects to several recent developments: higher-dimensional gravity on tori at null infinity, where the BMS group is extended by internal isometries~\cite{Ferko:2021bym}; the asymptotic symmetries of electromagnetism with the scalar couplings of extended supergravity at spatial infinity, where the boundary conditions select the compact subgroup of the duality group~\cite{Fuentealba:2024lll}; and the electric--magnetic duality of soft charges~\cite{Strominger:2015bla,Hosseinzadeh:2018dkh,Freidel:2018fsk}.

Three questions then arise naturally. How does S-duality act on the radiative data, on the balance laws and on the memory observables at null infinity, and what distinguishes S-dual processes there? Do mutually nonlocal charges, which are intrinsic to the duality-invariant theory, leave an imprint of their own on the memories? At which level of the asymptotic hierarchy do the curvature couplings fixed by the Green--Schwarz mechanism first appear explicitly? In this work we answer the three questions for the bosonic sector of the heterotic string on $T^6$, in the consistent truncation that keeps the four-dimensional metric, the axion--dilaton and the gauge fields from ten dimensions, through first order in $\alpha'$.

The first answer is a duality structure of the complete asymptotic problem. The scalar flux in the Bondi mass loss is the invariant metric of the moduli space $SL(2,\mathbb R)/U(1)$ evaluated on the scalar news, the gauge flux is the norm of a self-dual radiation field, and the balance laws, the charge evolution and the memories are covariant under $SL(2,\mathbb R)$ acting together with the vacuum value of $\tau$. At fixed vacuum the residual duality rotation, the null-infinity counterpart of the compact subgroup selected at spatial infinity, acts with charges zero, one and two on gravitational, electromagnetic and dilaton--axion radiation. S-dual processes therefore share their gravitational waveform and their gravitational memories, and are distinguished by their scalar memory, which for processes connecting stationary black holes is fixed by the charges and masses of the endpoints. The second answer is a dyonic contribution to the spin memory. Its ordinary part is governed by the Dirac--Schwinger--Zwanziger product of the total charge with the change of the charge aspect, it is generated in the scattering of mutually nonlocal black holes, and its dipole part agrees in magnitude and sign with the angular momentum radiated by a charge scattered off a magnetic monopole. {For a monopole embedded in one $E_8$ factor, whose charge is a topological integer of the reduction $E_8\to(E_7\times U(1))/\mathbb Z_2$ on the celestial sphere, it is the product of the quantized magnetic charge with the change of the electric charge aspect.} The third answer locates the first curvature correction. Within the four-derivative effective action the Gauss--Bonnet and Pontryagin couplings leave the mass-loss and angular-momentum balance laws free of explicit corrections, and first enter the evolution of the spin-two curvature coefficient $\Psi^0_0$, where the dilaton and the axion combine into the complex scalar charge aspect and the residual duality rotation is broken. Since the scalar charge of a black hole with fixed physical charges is of order one in the inverse string tension, this is a first-order effect in $\alpha'/M^2$, in contrast with scalar--Gauss--Bonnet and dynamical Chern--Simons models without a gauge sector.

The paper is organized as follows. Section~\ref{sec:action} presents the effective action and the consistent truncation, including the four-derivative couplings and the embedding of the abelian sector in the ten-dimensional gauge group. Section~\ref{sec:asymptotics} constructs the asymptotic solution space, and Sec.~\ref{sec:balance} derives the balance laws. Section~\ref{sec:duality} establishes the action of S-duality at null infinity. Section~\ref{sec:memory} treats the displacement, scalar and electromagnetic memories, including the scalar memory of black-hole processes, and Sec.~\ref{sec:spin} the spin memory and its dyonic sources, including the scattering of mutually nonlocal black holes. Section~\ref{sec:alpha} derives the first curvature correction at null infinity, and Sec.~\ref{sec:conclusions} collects our conclusions and the directions this work opens. The consistency of the truncation, the classification of the four-derivative operators, the leading components of the fields at null infinity and the curvature tensors of the Gauss--Bonnet and Pontryagin couplings are presented in Appendices~\ref{app:truncation}--\ref{app:curvature}. All results were checked against an independent symbolic integration of the Bondi--Sachs equations on a metric carrying both polarizations of the shear. We use signature $(-,+,+,+)$ and units with $G=c=1$.

%=====================================================================
\section{Effective action and consistent truncation}
\label{sec:action}

The two-derivative bosonic action of heterotic string theory compactified on $T^6$ can be written in the Einstein frame in the $O(6,22)$-covariant form of Refs.~\cite{Maharana:1992my,Sen:1994fa}. It contains the metric $g_{\mu\nu}$, the four-dimensional dilaton $\phi$, the Kalb--Ramond field $B_{\mu\nu}$, a $28\times28$ matrix $M$ of moduli built from the internal metric $G_{mn}$, the internal two-form $B_{mn}$ and the Wilson lines $A^I_m$, and 28 abelian vectors coupled through the kinetic matrix $LML$.

We truncate to the sector with constant internal metric and two-form, vanishing Wilson lines, and vanishing Kaluza--Klein and winding vectors $G_{\mu m}$, $B_{\mu m}$. At vanishing Wilson lines the kinetic matrix of the vectors descending from the ten-dimensional gauge fields is the identity and does not depend on $G_{mn}$ or $B_{mn}$, it depends on the Wilson lines only quadratically, and its mixing with the Kaluza--Klein and winding vectors vanishes. As shown in Appendix~\ref{app:truncation}, the field equations of all discarded fields are then identically satisfied, $\delta S/\delta\Phi_{\rm discarded}|_{\Phi_{\rm discarded}=0}=0$, for arbitrary configurations of the retained fields. In ten-dimensional terms, the truncation keeps gauge fields with vanishing internal components and no dependence on the internal coordinates, and it is consistent for an arbitrary four-dimensional configuration of the full ten-dimensional gauge group. The mass-loss formula and the scalar sector below hold in this generality; the statements involving electric--magnetic duality, the charge aspect and the spin memory refer to the abelian Cartan sector, on which the duality group acts. Dualizing $B_{\mu\nu}$ into the axion $a$, the truncated two-derivative action is
\begin{equation}
S_0=\frac{1}{16\pi}\int d^4x\,\sqrt{-g}\,\Big[R-2(\nabla\phi)^2-\tfrac12e^{4\phi}(\nabla a)^2
-\tfrac{\alpha'}{8}\,e^{-2\phi}F^I_{\mu\nu}F^{I\mu\nu}-\tfrac{\alpha'}{8}\,a\,F^I_{\mu\nu}\tilde F^{I\mu\nu}\Big],
\label{eq:action}
\end{equation}
where $I$ runs over the adjoint of the gauge group with a positive-definite normalization of the trace and $\tilde F^{\mu\nu}=\tfrac12\epsilon^{\mu\nu\rho\sigma}F_{\rho\sigma}$, with $\epsilon_{ur\theta\varphi}=+\sqrt{-g}$ in the coordinates introduced below. In terms of $\tau\equiv a+ie^{-2\phi}$,
\begin{equation}
-2(\nabla\phi)^2-\tfrac12e^{4\phi}(\nabla a)^2=-\frac{\nabla_\mu\tau\,\nabla^\mu\bar\tau}{2\,(\im\tau)^2},
\qquad
-\tfrac{\alpha'}{8}\big(e^{-2\phi}F\cdot F+a\,F\cdot\tilde F\big)=-\tfrac{\alpha'}{8}\,\im\big[\tau\,(F\cdot F+iF\cdot\tilde F)\big],
\label{eq:sigma}
\end{equation}
so that the scalar kinetic term is the invariant metric on $SL(2,\mathbb R)/U(1)$. The abelian sector of interest below is the Cartan torus of $E_8\times E_8$ or $Spin(32)/\mathbb Z_2$. {For definiteness, consider the $E_8\times E_8$ theory and a configuration supported in one $E_8$ factor. A magnetic configuration along a Cartan generator $\mathsf T$ whose centralizer is $\mathfrak e_7\oplus\mathfrak u(1)$ breaks $E_8$ to $H=(E_7\times U(1))/\mathbb Z_2$, and its Coulomb field is described by the abelian field of Eq.~\eqref{eq:action} along $\mathsf T$, with leading magnetic field $F^{(0)}_{AB}=p\,\mathsf T\,\epsilon_{AB}$ on the celestial sphere. Globally, the reduction of the structure group on the sphere is specified by a map $\varphi:S^2\to E_8/H$ into the adjoint orbit of $\mathsf T$, a homogeneous space of dimension $248-134=114$, and the $H$ bundle is the pullback of $H\to E_8\to E_8/H$ by $\varphi$. Since $\pi_1(E_8)=\pi_2(E_8)=0$, the exact homotopy sequence of this fibration gives $\pi_2(E_8/H)\simeq\pi_1(H)\simeq\mathbb Z$. The $E_8$ bundle over the sphere is therefore trivial, while its reduction to $H$ carries an integer, the first Chern class of the $U(1)$ factor, which is the monopole number. In the notation of Sec.~\ref{sec:asymptotics}, $p$ is the monopole of the magnetic charge aspect $f$ along $\mathsf T$, and its quantization is the Dirac--Schwinger--Zwanziger condition of Sec.~\ref{sec:duality}. The resulting contribution to the spin memory is given in Sec.~\ref{sec:spin}.}

The factor $\alpha'$ in front of the gauge kinetic terms is the normalization of the heterotic gauge fields inherited from ten dimensions, not a higher-derivative correction. In terms of the canonically normalized fields $\hat A^I_\mu=(\alpha'/8)^{1/2}A^I_\mu$, the two-derivative action is $\alpha'$-independent, and every two-derivative result below depends on $\alpha'$ only through the combinations $\alpha'\Qc^2$, $\alpha'\bar\Qc\psi_A$ and $\alpha'|\psi|^2$, which are $\alpha'$-independent when expressed through the canonical charges and radiation fields. We keep the heterotic normalization for comparison with the string literature and state the canonical form where it matters, in Sec.~\ref{sec:memory}. The genuine higher-derivative corrections are suppressed by $\alpha'/L^2$, with $L$ the curvature scale of the sources or the wavelength of the radiation, and we work to first order in this ratio.

At first order in $\alpha'$ the heterotic effective action contains, besides the gauge kinetic terms already present in Eq.~\eqref{eq:action}, four-derivative curvature terms fixed by the Green--Schwarz mechanism~\cite{Green:1984sg,Gross:1986mw,Metsaev:1987zx,Bergshoeff:1989de}. They enter through the combination ${\rm Tr}F^2-{\rm tr}R^2$ in the action and through the Lorentz Chern--Simons term in the field strength of the Kalb--Ramond field, whose Bianchi identity becomes $dH\propto{\rm tr}R\wedge R-{\rm Tr}F\wedge F$. After reduction, dualization and passage to the Einstein frame, the axion couples to the Pontryagin density with a coefficient fixed relative to its coupling to $F\tilde F$~\cite{Campbell:1990fu}, and the curvature-squared terms reduce to the Gauss--Bonnet density multiplied by $e^{-2\phi}$~\cite{Zwiebach:1985uq,Metsaev:1987zx,Kanti:1995vq,Cano:2019ycn}:
\begin{equation}
S_{\alpha'}=\frac{1}{16\pi}\int d^4x\sqrt{-g}\;\frac{\alpha'}{8}\Big(e^{-2\phi}\,\GB+a\,\Po\Big)
=\frac{1}{16\pi}\int d^4x\sqrt{-g}\;\frac{\alpha'}{8}\,\im\big[\tau\,(\GB+i\Po)\big],
\label{eq:Salpha}
\end{equation}
with
\begin{equation}
\GB=R_{\mu\nu\rho\sigma}R^{\mu\nu\rho\sigma}-4R_{\mu\nu}R^{\mu\nu}+R^2,\qquad
\Po=R_{\mu\nu\rho\sigma}\tilde R^{\mu\nu\rho\sigma},\qquad \tilde R^{\mu\nu\rho\sigma}=\tfrac12\epsilon^{\mu\nu\alpha\beta}R_{\alpha\beta}{}^{\rho\sigma}.
\end{equation}
The curvature couplings mirror the gauge couplings of Eq.~\eqref{eq:sigma}, with the opposite overall sign dictated by ${\rm Tr}F^2-{\rm tr}R^2$. The constant part $\tau_0$ multiplies topological densities, so that Eq.~\eqref{eq:Salpha} is invariant under the shifts $\tau\to\tau+b$ but not under the full $SL(2,\mathbb R)$.

In the field-redefinition scheme of Bergshoeff and de Roo~\cite{Bergshoeff:1989de}, specified in Appendix~\ref{app:operators}, we organize the four-dimensional effective action as
\begin{equation}
S_{\rm eff}=S_0+S_{\alpha'}+S_{\rm other}+O(\alpha'^2),
\label{eq:Seff}
\end{equation}
where $S_{\rm other}$ collects the remaining four-derivative operators generated by the reduction, the dualization and the passage to the Einstein frame. They fall into two classes, analyzed in Appendix~\ref{app:operators}: operators proportional to the Ricci tensor or to the lowest-order field equations, which are removed by field redefinitions $g_{\mu\nu}\to g_{\mu\nu}+O(\alpha')$ quadratic in the matter fields, and operators at least quartic in the first derivatives of $\tau$, which include the terms quartic in the Kalb--Ramond field strength. On the asymptotic solutions of Sec.~\ref{sec:asymptotics} the field redefinitions change the metric only at orders below those that carry the mass aspect, the angular-momentum aspect and the trace-free $O(r^{-3})$ angular coefficient $e_{AB}$, and the quartic operators produce stress tensors decaying at least as $r^{-5}$. In particular, the trace-free part of $\delta S_{\rm other}/\delta g^{AB}$ does not enter the angular equation at the order where $e_{AB}$ evolves, and $S_{\rm other}$ affects none of the results below. The truncation of Appendix~\ref{app:truncation} remains consistent in the presence of $S_{\alpha'}$ and $S_{\rm other}$.

The field equations of $S_0+S_{\alpha'}$ are
\begin{gather}
G_{\mu\nu}=8\pi T_{\mu\nu}-\frac{\alpha'}{2}\,H_{\mu\nu}\big[e^{-2\phi}\big]+\frac{\alpha'}{2}\,C_{\mu\nu}[a],
\label{eq:einstein}\\
\Box\phi=\tfrac12e^{4\phi}(\nabla a)^2-\tfrac{\alpha'}{16}e^{-2\phi}\big(F^I_{\mu\nu}F^{I\mu\nu}-\GB\big),\qquad
\nabla_\mu\big(e^{4\phi}\nabla^\mu a\big)=\tfrac{\alpha'}{8}\big(F^I_{\mu\nu}\tilde F^{I\mu\nu}-\Po\big),
\label{eq:scalars}\\
\nabla_\mu\big(e^{-2\phi}F^{\mu\nu}+a\tilde F^{\mu\nu}\big)=0,\qquad \nabla_\mu\tilde F^{\mu\nu}=0
\label{eq:maxwell}
\end{gather}
(the Maxwell equations are written for each abelian factor; in the non-abelian case they are replaced by their gauge-covariant version). The matter stress tensor is
\begin{equation}
8\pi T_{\mu\nu}=2\Big(\nabla_\mu\phi\nabla_\nu\phi-\tfrac12g_{\mu\nu}(\nabla\phi)^2\Big)
+\tfrac12e^{4\phi}\Big(\nabla_\mu a\nabla_\nu a-\tfrac12g_{\mu\nu}(\nabla a)^2\Big)
+\tfrac{\alpha'}{4}e^{-2\phi}\Big(F^I_{\mu\rho}F^{I}{}_\nu{}^\rho-\tfrac14g_{\mu\nu}F^I_{\rho\sigma}F^{I\rho\sigma}\Big),
\label{eq:T}
\end{equation}
and the curvature couplings contribute through
\begin{gather}
H_{\mu\nu}[f]=P_{\mu\alpha\nu\beta}\nabla^\alpha\nabla^\beta f,\qquad
P_{\mu\alpha\nu\beta}=R_{\mu\alpha\nu\beta}-2g_{\mu[\nu}R_{\beta]\alpha}+2g_{\alpha[\nu}R_{\beta]\mu}+R\,g_{\mu[\nu}g_{\beta]\alpha},
\label{eq:H}\\
C^{\mu\nu}[\vartheta]=\nabla_\sigma\vartheta\,\epsilon^{\sigma\alpha\beta(\mu}\nabla_\beta R^{\nu)}{}_\alpha+\nabla_\sigma\nabla_\tau\vartheta\,{}^\star R^{\tau(\mu\nu)\sigma},\qquad
{}^\star R^{\tau\mu\nu\sigma}=\tfrac12\epsilon^{\nu\sigma\alpha\beta}R^{\tau\mu}{}_{\alpha\beta},
\label{eq:C}
\end{gather}
where $P$ is the double dual of the Riemann tensor and $C$ is the Cotton-like tensor of the Pontryagin coupling~\cite{Jackiw:2003pm,Alexander:2009tp}, with $\epsilon_{ur\theta\varphi}=+\sqrt{-g}$, $\epsilon^{ur\theta\varphi}=-1/\sqrt{-g}$. The metric variations of $\int\sqrt{-g}\,f\GB$ and $\int\sqrt{-g}\,\vartheta\Po$ are proportional to $H_{\mu\nu}[f]$ and $C_{\mu\nu}[\vartheta]$, the unique symmetric tensors with this field content and derivative order~\cite{Horndeski:1974wa,Jackiw:2003pm}, and their coefficients in Eq.~\eqref{eq:einstein} follow from the Noether identity of diffeomorphism invariance together with
\begin{equation}
\nabla^\mu H_{\mu\nu}[f]=-\tfrac18\,\GB\,\nabla_\nu f,\qquad \nabla_\mu C^{\mu\nu}[\vartheta]=\tfrac18\,\Po\,\nabla^\nu\vartheta,
\label{eq:HCidentities}
\end{equation}
which guarantee that the divergence of Eq.~\eqref{eq:einstein} vanishes on the scalar equations~\eqref{eq:scalars}. The derivation of Eqs.~\eqref{eq:HCidentities} and of the coefficients is given in Appendix~\ref{app:curvature}. The two scalar equations combine into the $SL(2,\mathbb R)$-covariant equation $\Box\tau+i(\nabla\tau)^2/\im\tau=\mathcal S$, where $\mathcal S$ is quadratic in the field strengths and in the curvature.

%=====================================================================
\section{Asymptotic solution space}
\label{sec:asymptotics}

We use retarded Bondi coordinates $(u,r,x^A)$ and the Bondi--Sachs metric~\cite{Bondi:1962px,Sachs:1962wk,Madler:2016xju}
\begin{equation}
ds^2=-\frac Vr\,e^{2\beta}du^2-2e^{2\beta}du\,dr+r^2h_{AB}\big(dx^A-U^Adu\big)\big(dx^B-U^Bdu\big),\qquad \det h_{AB}=\det q_{AB},
\label{eq:bondi}
\end{equation}
where $q_{AB}$ is the unit-sphere metric with covariant derivative $D_A$ and volume form $\epsilon_{AB}$, $\epsilon_{\theta\varphi}=\sin\theta$. Asymptotic flatness is imposed through
\begin{gather}
h_{AB}=q_{AB}+\frac{c_{AB}}{r}+\frac{q_{AB}\,c_{CD}c^{CD}}{4r^2}+\frac{e_{AB}}{r^3}+O(r^{-4}),\qquad
\beta=\frac{\beta_2}{r^2}+O(r^{-3}),\notag\\
U^A=-\frac{D_Bc^{AB}}{2r^2}+O(r^{-3}),\qquad V=r-2M+O(r^{-1}),
\label{eq:metricfalloff}
\end{gather}
with $c_{AB}(u,x^C)$ the trace-free shear, $N_{AB}=\partial_uc_{AB}$ the Bondi news and $M(u,x^C)$ the mass aspect. For the matter fields we take
\begin{gather}
\tau=\tau_0+\frac{\tau_1(u,x^C)}{r}+\frac{\tau_2(u,x^C)}{r^2}+O(r^{-3}),\notag\\
A^I_u=\frac{Q^I(u,x^C)}{r}+O(r^{-2}),\qquad A^I_r=0,\qquad A^I_A=A^{I(0)}_A(u,x^C)+O(r^{-1}),
\label{eq:matterfalloff}
\end{gather}
in the radial gauge for the gauge fields, with
\begin{equation}
\tau_1=a_1-2ie^{-2\phi_0}\,\phi_1,\qquad \im\tau_0=e^{-2\phi_0}.
\end{equation}
For sectors with magnetic charge, the angular potential $A^{I(0)}_A$ is defined patchwise on the sphere with time-independent transition functions, whereas the field strength, the radiation field and all fluxes are global. The leading field strengths are $F_{ur}=Q/r^2$, $F_{uA}=\dot A^{(0)}_A$ and $F_{AB}=f\,\epsilon_{AB}$, where a dot denotes $\partial_u$ and $f$ is the magnetic charge aspect, $F^{(0)}_{AB}=f\epsilon_{AB}$. The vacuum value $\tau_0$ is constant: the $O(r^0)$ part of the $uu$ component of Eq.~\eqref{eq:einstein} contains $2(\partial_u\phi_0)^2+\tfrac12e^{4\phi_0}(\partial_ua_0)^2$, a sum of non-negative terms that must vanish, and an angle-dependent $\tau_0$ would produce a non-decaying trace-free part of $T_{AB}$.

Two expansions are involved and must be kept distinct: the asymptotic expansion in $1/r$ and the effective-field-theory expansion in $\alpha'$. Every field is a solution of the full equations derived from $S_{\rm eff}$, $X=X^{(0)}+\alpha'X^{(1)}+O(\alpha'^2)$, and each $X^{(n)}$ admits the smooth expansion $\sum_kX^{(n)}_k(u,x^A)\,r^{-k}$. The coefficients $c_{AB}$, $M$, $N_A$, $\tau_1$, $\Qc$, $e_{AB}$ appearing below are the complete coefficients, including their own $O(\alpha')$ parts, and each equation of the hierarchy holds order by order in $1/r$ through first order in $\alpha'$. When we state that a curvature coupling does not enter an equation, we mean that it contributes no explicit term to that equation at the given order in $1/r$; the data in the equation may themselves differ from their two-derivative values at $O(\alpha')$. The determinant condition $\det h_{AB}=\det q_{AB}$ holds exactly in the parametrization $h_{AB}dx^Adx^B=\cosh2\delta\,\big(e^{2\gamma}d\theta^2+e^{-2\gamma}\sin^2\theta\,d\varphi^2\big)+2\sinh2\delta\,\sin\theta\,d\theta\,d\varphi$, in which $\gamma$ and $\delta$ carry the two polarizations of the shear.

Inserting the expansions into Eq.~\eqref{eq:T}, the components of the stress tensor decay as
\begin{gather}
T_{rr}=O(r^{-4}),\qquad T_{rA}=O(r^{-3}),\qquad T_{ur}=O(r^{-4}),\notag\\
T^{\rm TF}_{AB}=O(r^{-2}),\qquad h^{AB}T_{AB}=O(r^{-1}),\qquad T_{uu},\,T_{uA}=O(r^{-2}).
\label{eq:Tfalloff}
\end{gather}
For the scalars, the $O(r^{-3})$ parts of $\partial_u\phi\,\partial_r\phi$ and of $\tfrac12(\nabla\phi)^2$ in $T_{ur}$ cancel, and the $O(r^{-1})$ part of $T_{AB}$ is pure trace and equal to $-\tfrac{1}{8\pi}q_{AB}\,\partial_u\big(\phi_1^2+\tfrac14e^{4\phi_0}a_1^2\big)$. For the gauge fields $F^2=O(r^{-4})$, so that the gauge part of every component is subleading except for $T_{uu}$ and $T_{uA}$. These falloffs satisfy all the conditions derived in Ref.~\cite{Maibach:2026afb} for an arbitrary stress tensor sourcing the Bondi--Sachs hierarchy: $T^{(3)}_{rr}=T^{(2)}_{rA}=T^{(3)}_{ur}=0$ and $T^{(1)}_{AB}$ is pure trace. The hypersurface equations then integrate to the expansion~\eqref{eq:metricfalloff}. The $O(r^{-3})$ coefficient of the $rA$ equations, which controls the appearance of $\log r$ terms in $U^A$, contains no matter contribution, and the absence of logarithms requires, as in vacuum, that the trace-free $O(r^{-2})$ coefficient of $h_{AB}$ vanish for both polarizations, as anticipated in Eq.~\eqref{eq:metricfalloff}. The leading coefficient of $\beta$ is
\begin{equation}
\beta_2=-\frac{1}{32}c_{AB}c^{AB}-\frac{|\tau_1|^2}{16\,(\im\tau_0)^2},
\label{eq:beta2}
\end{equation}
where the scalar term is again the moduli-space metric. The theory is minimally coupled in the Einstein frame at two derivatives, so the asymptotic Killing analysis is that of vacuum general relativity and the asymptotic symmetry group of the metric sector is the BMS group, extended by superrotations in the generalized sense~\cite{Barnich:2009se,Kapec:2014opa,Campiglia:2014yka,Flanagan:2015pxa}.

For the scalars, the d'Alembertian on the Bondi--Sachs background reads
\begin{equation}
\Box\phi=\frac{2\,\partial_u\phi_2+D^2\phi_1}{r^3}+O(r^{-4}),
\label{eq:box}
\end{equation}
and analogously for $a$. The $O(r^{-2})$ terms $-2\partial_u\partial_r\phi$ and $-\tfrac2r\partial_u\phi$ cancel identically, so the field equations impose no condition on $\partial_u\phi_1$ or $\partial_ua_1$: the coefficients $\phi_1(u,x^A)$ and $a_1(u,x^A)$, equivalently $\tau_1(u,x^A)$, are free radiative data on the same footing as the shear $c_{AB}$. The gauge-field sources in Eq.~\eqref{eq:scalars} are $O(r^{-4})$, the curvature densities $\GB$ and $\Po$ decay as $r^{-6}$ (Sec.~\ref{sec:alpha}), and the axion--dilaton cross terms start at $O(r^{-3})$, with $(\nabla a)^2=2a_1\dot a_1/r^3+O(r^{-4})$ and $\nabla\phi\cdot\nabla a=\partial_u(\phi_1a_1)/r^3+O(r^{-4})$. At $O(r^{-3})$ the scalar equations give
\begin{equation}
2\,\partial_u\phi_2+D^2\phi_1=e^{4\phi_0}a_1\dot a_1,\qquad
2\,\partial_ua_2+D^2a_1=-4\,\partial_u(\phi_1a_1),
\label{eq:scalarhier}
\end{equation}
which combine into the single $SL(2,\mathbb R)$-covariant equation
\begin{equation}
2\,\partial_u\tau_2+D^2\tau_1+\frac{i}{\im\tau_0}\,\partial_u\big(\tau_1^2\big)=0 .
\label{eq:tauhier}
\end{equation}
The gauge fields first enter the scalar hierarchy in the evolution of $\tau_3$, and the curvature couplings in that of $\tau_5$.

For the abelian gauge fields, the $O(r^{-2})$ component of the Maxwell equation along $\nu=r$ and the Bianchi identity give
\begin{equation}
\partial_uQ=D^A\dot A^{(0)}_A,\qquad \partial_uf=\epsilon^{AB}D_A\dot A^{(0)}_B .
\label{eq:gauss}
\end{equation}
The axion contributes to Eq.~\eqref{eq:maxwell} through $\nabla_\mu a\,\tilde F^{\mu\nu}$, which is $O(r^{-3})$ and does not enter at this order. In the non-abelian case the derivatives are gauge covariant and the first equation acquires the color current of the radiated gauge bosons.

%=====================================================================
\section{Balance laws}
\label{sec:balance}

The supplementary $uu$ component of the field equations at $O(r^{-2})$ gives, for an arbitrary stress tensor obeying Eq.~\eqref{eq:Tfalloff}~\cite{Maibach:2026afb,Flanagan:2015pxa},
\begin{equation}
\partial_uM=\tfrac14D_AD_BN^{AB}-\tfrac18N_{AB}N^{AB}-4\pi T^{(2)}_{uu},\qquad
8\pi T^{(2)}_{uu}=2\dot\phi_1^{\,2}+\tfrac12e^{4\phi_0}\dot a_1^{\,2}+\tfrac{\alpha'}{4}e^{-2\phi_0}\,q^{AB}\dot A^{I(0)}_A\dot A^{I(0)}_B .
\end{equation}
To exhibit the structure of the gauge flux we introduce, for each abelian factor, the dual field strength and the complex combination
\begin{equation}
G_{\mu\nu}\equiv\im\tau\;{}^\star F_{\mu\nu}-\re\tau\;F_{\mu\nu},\qquad \Psi_{\mu\nu}\equiv G_{\mu\nu}+\tau F_{\mu\nu}=\im\tau\;{}^\star F_{\mu\nu}+i\,\im\tau\;F_{\mu\nu},
\label{eq:Psi}
\end{equation}
with ${}^\star F_{\mu\nu}=\tfrac12\epsilon_{\mu\nu\rho\sigma}F^{\rho\sigma}$ and $\epsilon_{ur\theta\varphi}=+\sqrt{-g}$, for which ${}^\star{}^\star F=-F$ in Lorentzian signature. The Maxwell equation~\eqref{eq:maxwell} is $dG=0$, and $\Psi$ is self-dual for every value of $\tau$, ${}^\star\Psi=\im\tau\,({}^\star{}^\star F+i\,{}^\star F)=i\Psi$. At null infinity ${}^\star F^{(2)}_{ur}=f$ and ${}^\star F^{(0)}_{uA}=\epsilon_A{}^B\dot A^{(0)}_B$, so the leading components of $\Psi$ define the complex charge aspect and the complex radiation field
\begin{equation}
\Qc\equiv\Psi^{(2)}_{ur}=\im\tau_0\,f+i\,\im\tau_0\,Q,\qquad
\psi_A\equiv\Psi^{(0)}_{uA}=\im\tau_0\;\epsilon_A{}^B\dot A^{(0)}_B+i\,\im\tau_0\;\dot A^{(0)}_A,\qquad
\epsilon_A{}^B\psi_B=i\psi_A .
\label{eq:psi}
\end{equation}
The self-duality condition leaves one complex function, i.e.\ the two radiative polarizations of the gauge field. These relations, the self-duality of $\Psi$ and the identities below hold for arbitrary $a_0=\re\tau_0$ and $\phi_0$. Since $q^{AB}\psi_A\bar\psi_B=2(\im\tau_0)^2q^{AB}\dot A^{(0)}_A\dot A^{(0)}_B$, the mass-loss formula becomes
\begin{equation}
{\;\partial_uM=\tfrac14D_AD_BN^{AB}-\tfrac18N_{AB}N^{AB}-\frac{|\partial_u\tau_1|^2}{4\,(\im\tau_0)^2}-\frac{\alpha'}{16}\,\frac{q^{AB}\psi_A\bar\psi_B}{\im\tau_0}\;}
\label{eq:massloss}
\end{equation}
with an implicit sum over the abelian factors in the last term; for non-abelian fields the last term is $\tfrac{\alpha'}{8}e^{-2\phi_0}q^{AB}\dot A^{I(0)}_A\dot A^{I(0)}_B$ with the trace over the adjoint. Each of the three fluxes is non-negative. The scalar flux is the line element of the upper half-plane at $\tau_0$ evaluated on the scalar news; the dilaton contributes through $\dot\phi_1^{\,2}$ at the same order as the axion and the gauge fields.

The supplementary $uA$ components at $O(r^{-2})$ govern the angular-momentum aspect $N_A$, which we define through~\cite{Flanagan:2015pxa}
\begin{equation}
g_{uA}=\tfrac12D^Bc_{AB}+\frac{2}{3r}\Big(N_A+\tfrac14c_{AB}D_Cc^{BC}\Big)+O(r^{-2}).
\end{equation}
For the truncated theory we find
\begin{equation}
{\;
\begin{aligned}
\partial_uN_A={}&D_AM+\tfrac14D_BD_AD_Cc^{BC}-\tfrac14D^2D^Bc_{AB}+\mathcal J_A
-8\pi T^{(2)}_{uA}
+\partial_uD_A\Big(\frac{|\tau_1|^2}{16(\im\tau_0)^2}-\frac{3}{32}c_{BC}c^{BC}\Big)
\end{aligned}\;}
\label{eq:Nevol}
\end{equation}
with the gravitational term
\begin{equation}
\mathcal J_A=\tfrac14D_B\big(N^{BC}c_{AC}\big)+\tfrac14c_{AC}D_BN^{BC}-\tfrac14N_{AC}D_Bc^{BC},
\label{eq:JA}
\end{equation}
and the matter flux
\begin{equation}
8\pi T^{(2)}_{uA}=\frac{\re\big(\partial_u\tau_1\,D_A\bar\tau_1\big)}{2(\im\tau_0)^2}+\frac{\alpha'}{4}\,\frac{\re\big(\bar\Qc\,\psi_A\big)}{\im\tau_0}.
\label{eq:TuA}
\end{equation}
The gauge term equals $\tfrac{\alpha'}{4}e^{-2\phi_0}\big(Q\dot A^{(0)}_A+\dot A^{(0)C}F^{(0)}_{AC}\big)$, the sum of the cross terms between the radiation and the electric and magnetic Coulomb fields. The scalar terms are $2\dot\phi_1D_A\phi_1+\tfrac12e^{4\phi_0}\dot a_1D_Aa_1$. The last term of Eq.~\eqref{eq:Nevol} is a total gradient: it does not contribute to the magnetic-parity projection of the balance law and can be absorbed into a redefinition of $N_A$. The scalar part of it is the trace of $T_{rr}$ at $O(r^{-4})$ entering the hypersurface equation for $U^A$, and it is again the moduli-space metric.

In terms of the complex variables~\eqref{eq:psi}, the two equations~\eqref{eq:gauss} combine into
\begin{equation}
\partial_u\Qc=D^A\psi_A .
\label{eq:Qevol}
\end{equation}
The electric and magnetic charge aspects evolve together, sourced by the divergence of the complex radiation field. Integrated over the sphere, Eq.~\eqref{eq:Qevol} expresses the conservation of the total electric and magnetic charges, since no massless state of the truncation is charged under the abelian factors.

%=====================================================================
\section{S-duality at null infinity}
\label{sec:duality}

For abelian gauge fields the two-derivative equations of motion are invariant under~\cite{Schwarz:1992tn,Sen:1994fa,Gibbons:1995ap}
\begin{equation}
\tau\to\frac{a\tau+b}{c\tau+d},\qquad
\begin{pmatrix}G\\F\end{pmatrix}\to\begin{pmatrix}a&-b\\-c&d\end{pmatrix}\begin{pmatrix}G\\F\end{pmatrix},\qquad ad-bc=1,
\label{eq:SL2}
\end{equation}
with the Einstein-frame metric invariant. This transformation preserves the constitutive relation~\eqref{eq:Psi} and acts on the self-dual field as $\Psi\to\Psi/(c\tau+d)$, while $\im\tau\to\im\tau/|c\tau+d|^2$. Expanding at null infinity, the asymptotic data transform as
\begin{gather}
\tau_0\to\frac{a\tau_0+b}{c\tau_0+d},\qquad
\tau_1\to\frac{\tau_1}{(c\tau_0+d)^2},\qquad
\tau_2\to\frac{\tau_2}{(c\tau_0+d)^2}-\frac{c\,\tau_1^2}{(c\tau_0+d)^3},\notag\\
\psi_A\to\frac{\psi_A}{c\tau_0+d},\qquad \Qc\to\frac{\Qc}{c\tau_0+d},\qquad
c_{AB}\to c_{AB},\qquad M\to M,\qquad N_A\to N_A .
\label{eq:SL2data}
\end{gather}
The following properties follow directly. The scalar and gauge fluxes in Eqs.~\eqref{eq:massloss} and~\eqref{eq:TuA} are separately invariant, and so is the gradient term in Eq.~\eqref{eq:Nevol}; the mass loss and the angular-momentum balance are therefore S-duality invariant. The scalar hierarchy~\eqref{eq:tauhier} and the charge evolution~\eqref{eq:Qevol} are covariant; in Eq.~\eqref{eq:tauhier} the inhomogeneous transformation of $\tau_2$ is compensated by the nonlinear term $i\partial_u(\tau_1^2)/\im\tau_0$, as follows from $i(\bar z-z)/\im\tau_0=2c$ with $z=c\tau_0+d$. The quantity
\begin{equation}
\langle\Qc_1,\Qc_2\rangle\equiv\frac{\im\big(\bar\Qc_1\Qc_2\big)}{\im\tau_0}
\label{eq:DSZ}
\end{equation}
is invariant under the full group~\eqref{eq:SL2}. In terms of the magnetic charge aspect $f$ and of the electric charge aspect conjugate to the Maxwell equation $dG=0$,
\begin{equation}
\mathsf q_e\equiv\im\tau_0\,Q-\re\tau_0\,f,\qquad \Qc=i\,\mathsf q_e+i\,\bar\tau_0\,f,
\label{eq:qe}
\end{equation}
that is, $\re\Qc=\im\tau_0\,f$ and $\im\Qc=\mathsf q_e+\re\tau_0\,f=\im\tau_0\,Q$, in agreement with Eq.~\eqref{eq:psi} for every $\tau_0$. One then finds $\langle\Qc_1,\Qc_2\rangle=f_1\mathsf q_{e2}-\mathsf q_{e1}f_2$. This is the symplectic product of the electric and magnetic charges that enters the Dirac--Schwinger--Zwanziger quantization condition~\cite{Dirac:1931kp,Schwinger:1966nj,Zwanziger:1968rs}, and $(\mathsf q_e,f)$ transforms as a doublet under Eq.~\eqref{eq:SL2}. It plays a central role in Sec.~\ref{sec:spin}.

The transformation~\eqref{eq:SL2data} moves the vacuum value $\tau_0$. Since $\tau_0$ labels the asymptotic phase space, the subgroup that acts within a given phase space is the stabilizer of $\tau_0$, a $U(1)$ conjugate to $SO(2)$. This is the null-infinity counterpart of the breaking to the compact subgroup found at spatial infinity in Ref.~\cite{Fuentealba:2024lll}. On the stabilizer $c\tau_0+d=e^{-i\theta}$, and Eq.~\eqref{eq:SL2data} reduces to
\begin{equation}
\psi_A\to e^{i\theta}\psi_A,\qquad \Qc\to e^{i\theta}\Qc,\qquad \tau_1\to e^{2i\theta}\tau_1,\qquad c_{AB}\to c_{AB}.
\label{eq:U1}
\end{equation}
The residual duality rotation acts on the radiative data at null infinity with charge zero on the gravitational news, charge one on the electromagnetic news and charge two on the dilaton--axion news. It rotates electric into magnetic radiation and dilaton into axion radiation, leaving all fluxes invariant. The full $SL(2,\mathbb R)$ relates the asymptotic phase spaces with different $\tau_0$ and maps the memory observables of one into those of the other. Table~\ref{tab:data} collects the asymptotic data, their transformation under the residual rotation and the observables they control; the last row anticipates the result of Sec.~\ref{sec:alpha}.

\begin{table}[t]
\caption{Asymptotic data of the heterotic truncation at null infinity, their charge under the residual duality rotation~\eqref{eq:U1} and the balance laws and memory observables they enter.}
\label{tab:data}
\begin{ruledtabular}
\begin{tabular}{p{2.3cm}p{4.3cm}cp{6.3cm}}
Datum & Meaning & $U(1)$ charge & Enters \\
\hline
$c_{AB}$, $N_{AB}$ & shear, news & 0 & mass loss, $N_A$ balance, displacement and spin memory\\
$M$, $N_A$ & mass and angular-momentum aspects & 0 & Eqs.~\eqref{eq:massloss}, \eqref{eq:Nevol} \\
$\tau_1$ & dilaton--axion charge aspect & 2 & scalar flux, scalar memory $\Delta\tau_1$ \\
$\psi_A$ & self-dual radiation field & 1 & gauge flux, electromagnetic memory \\
$\Qc$ & complex charge aspect & 1 & Eq.~\eqref{eq:Qevol}, dyonic spin memory via $\langle\Qc_0,\Delta\Qc\rangle$ \\
$\Psi_0^0$ & spin-two aspect & 0 & $O(\alpha')$ source $\tfrac{3i\alpha'}{2}\tau_1\bar\Psi_4^0$, of charge 2 \\
\end{tabular}
\end{ruledtabular}
\end{table}

The Witten effect~\cite{Witten:1979ey} is built into these variables: a shift $\tau_0\to\tau_0+b$ leaves $\Qc$, $\psi_A$ and the field aspects $Q$ and $f$ invariant, while it shifts the electric charge~\eqref{eq:qe} of a magnetic configuration by $-b\,f$.

%=====================================================================
\section{Displacement, scalar and electromagnetic memory}
\label{sec:memory}

A congruence of freely falling test masses at large $r$, initially at rest, obeys the geodesic deviation equation. The dominant part of the Riemann tensor is $R_{uAuB}=-\tfrac r2\,\partial_u^2c_{AB}+O(1)$, so that in an orthonormal frame on the sphere the permanent displacement after the passage of radiation is
\begin{equation}
\Delta\xi^{\hat A}=\frac{1}{2r}\,\Delta c^{\hat A}{}_{\hat B}\,\xi^{\hat B},\qquad \Delta c_{AB}=\int_{-\infty}^{\infty}du\,N_{AB}.
\end{equation}
Every trace-free tensor on the sphere decomposes into electric- and magnetic-parity parts,
\begin{equation}
c_{AB}=-\big(2D_AD_B-q_{AB}D^2\big)\Cc+2\,\epsilon_{(A}{}^CD_{B)}D_C\Ss,
\label{eq:EBdec}
\end{equation}
and in the nonradiative epochs before and after the burst the shear is a supertranslation of the vacuum, generated by $\delta_\alpha c_{AB}=\alpha N_{AB}-2D_AD_B\alpha+q_{AB}D^2\alpha$; it has electric parity, so that $\Delta c_{AB}=-(2D_AD_B-q_{AB}D^2)\Delta\Cc$. Since $D^AD^B(2D_AD_B-q_{AB}D^2)\Cc=D^2(D^2+2)\Cc$, integrating Eq.~\eqref{eq:massloss} in $u$ gives
\begin{equation}
{\;-\tfrac14D^2(D^2+2)\,\Delta\Cc=\Delta M+\int_{-\infty}^{\infty}du\Big[\tfrac18N_{AB}N^{AB}+\frac{|\partial_u\tau_1|^2}{4(\im\tau_0)^2}+\frac{\alpha'}{16}\frac{q^{AB}\psi_A\bar\psi_B}{\im\tau_0}\Big]\;}
\label{eq:memory}
\end{equation}
The operator $D^2(D^2+2)$ annihilates the $\ell=0,1$ harmonics, whose content is the conservation of the Bondi four-momentum, and is invertible on $\ell\geq2$ with a Green function $\mathcal G(\hat n,\hat n')$ on the sphere,
\begin{equation}
\Delta\Cc(\hat n)=-4\int d^2\Omega'\,\mathcal G(\hat n,\hat n')\Big\{\Delta M+\int du\big[\dots\big]\Big\}(\hat n').
\end{equation}
The first term on the right-hand side of Eq.~\eqref{eq:memory} is the ordinary memory and the integral is the null memory~\cite{Christodoulou:1991cr,Thorne:1992sdb,Strominger:2014pwa}. Compared with vacuum general relativity, the supertranslation $\Delta\Cc$ that connects the initial and final vacua is shifted by
\begin{equation}
\delta\Delta\Cc=-4\int d^2\Omega'\,\mathcal G(\hat n,\hat n')\int du\Big[\frac{|\partial_u\tau_1|^2}{4(\im\tau_0)^2}+\frac{\alpha'}{16}\frac{q^{AB}\psi_A\bar\psi_B}{\im\tau_0}\Big](\hat n'),
\label{eq:shift}
\end{equation}
whose source is non-negative and receives contributions from the dilaton, the axion and the gauge fields. It is S-duality invariant, so that the displacement memory of freely falling test masses cannot distinguish S-dual processes.

The matter sectors carry their own memory observables. The scalar memory is the net change $\Delta\tau_1$ of the asymptotic scalar charge aspect between the initial and final nonradiative epochs~\cite{Satishchandran:2019pyc,Hou:2020tnd,Tahura:2020vsa}. It is nonzero when the scalar charges of the sources change, e.g.\ through the formation of a remnant with different dilaton or axion charge, and it transforms with weight $-2$ under Eq.~\eqref{eq:SL2data}, i.e.\ with charge two under the residual rotation. The electromagnetic memory is the net change of the leading gauge potential~\cite{Bieri:2013hqa,Pasterski:2015zua}, which by Eq.~\eqref{eq:Qevol} satisfies
\begin{equation}
D^A\!\int du\,\psi_A=\Delta\Qc ,
\label{eq:EMmemory}
\end{equation}
and transforms with weight $-1$. It produces a net momentum kick on charged test particles, and in the truncation it is of the ordinary type, determined by the change in the complex charge aspect. For non-abelian fields the corresponding observable is the color memory of Ref.~\cite{Pate:2017vwa}.

The scalar memory becomes an endpoint relation for processes whose initial and final states are stationary black holes. Black holes are the natural sources of the theory: in the abelian truncation there is no adjoint scalar, so magnetic and dyonic charges are supported by horizons, and the stationary charged solutions are the magnetic black hole of Gibbons and Maeda and of Garfinkle, Horowitz and Strominger~\cite{Gibbons:1987ps,Garfinkle:1990qj} and its $SL(2,\mathbb R)$ images, which carry arbitrary electric and magnetic charges~\cite{Shapere:1991ta,Kallosh:1992ii,Kallosh:1993yh}; in the full string theory they describe the dyonic states of S-duality~\cite{Sen:1992ua,Sen:1994fa}. At null infinity a widely separated black hole enters through its mass, its complex charge and its scalar charge aspect, and on the stationary family the scalar charge is a secondary charge fixed by the other two. The relations below connect the asymptotic data of the initial state, widely separated stationary black holes, with those of the final state, a stationary remnant together with the radiation emitted. They follow from charge conservation and from the stationary family, and the radiated energy, which includes the gravitational, scalar and gauge radiation, enters only through the final mass. For the magnetic solution, in the normalization of Eq.~\eqref{eq:action}, with $F=P\sin\theta\,d\theta\wedge d\varphi$,
\begin{equation}
ds^2=-\Big(1-\frac{2M}{r}\Big)dt^2+\frac{dr^2}{1-2M/r}+r(r-\mathsf a)\,d\Omega^2,\qquad
e^{-2\phi}=e^{-2\phi_0}\Big(1-\frac{\mathsf a}{r}\Big),\qquad \mathsf a=\frac{\alpha'e^{-2\phi_0}P^2}{8M}.
\label{eq:GHS}
\end{equation}
Its asymptotic data are $\Qc=\im\tau_0\,P$, $a_1=0$ and $\phi_1=\mathsf a/2$, so that $\tau_1=-i\alpha'\Qc^2/(8M)$. The general static non-rotating black hole of the dilaton--axion theory is obtained from Eq.~\eqref{eq:GHS} by an $SL(2,\mathbb R)$ transformation~\cite{Shapere:1991ta,Kallosh:1992ii,Kallosh:1993yh}. Since $M$ is invariant while $\tau_1$ and $\Qc^2$ both have weight $-2$, the relation holds on the whole family:
\begin{equation}
\tau_1=-\frac{i\alpha'}{8}\,\frac{\Qc^2}{M}.
\label{eq:tau1BH}
\end{equation}
For an electric black hole $\Qc^2<0$ and the dilaton charge has the opposite sign to the magnetic case, while a dyon with $\Qc^2$ purely imaginary carries only axion charge. The relation extends to rotating black holes: for the Kerr--Sen solution~\cite{Sen:1992ua} the dilaton charge is independent of the rotation parameter and the axion field decays as $r^{-2}$, so that the monopole part of $\tau_1$ is again given by Eq.~\eqref{eq:tau1BH}, and the same holds for its $SL(2,\mathbb R)$ images.

The factor $\alpha'$ in Eq.~\eqref{eq:tau1BH} is the normalization of the heterotic gauge fields discussed in Sec.~\ref{sec:action}, and it does not make $\tau_1$ small. In terms of the canonically normalized complex charge $\hat\Qc\equiv(\alpha'/8\im\tau_0)^{1/2}\,\Qc$, which for the magnetic solution is the physical charge $\hat P=(\alpha'\im\tau_0/8)^{1/2}P$ measured at infinity in Einstein-frame units,
\begin{equation}
\tau_1=-i\,\im\tau_0\,\frac{\hat\Qc^2}{M},\qquad |\hat\Qc|^2\le2M^2,
\label{eq:tau1can}
\end{equation}
where the bound is the extremality condition $\mathsf a\le2M$ of Eq.~\eqref{eq:GHS} and its duality images. At fixed physical mass and charge, $\tau_1$ is of order one in $\alpha'$, $|\tau_1|/\im\tau_0\le2M$, and is comparable to the mass for near-extremal black holes. This is the scaling that governs the curvature correction of Sec.~\ref{sec:alpha}, whose expansion parameter is $\alpha'/M^2$.

Now consider a process in which $n$ such black holes, initially stationary and at large separations, with masses $m_i$ and charges $\Qc_i$, settle into a single stationary remnant of mass $M_f=\sum_im_i-E_{\rm rad}$, where $E_{\rm rad}$ is the total energy carried away by gravitational, scalar and gauge radiation. The total charge is conserved, $\Qc_f=\sum_i\Qc_i$. The monopole part of $\tau_1$ is additive at large separations, and Eq.~\eqref{eq:tau1BH} gives the $\ell=0$ scalar memory
\begin{equation}
{\;\Delta\tau_1^{(\ell=0)}=-\frac{i\alpha'}{8}\bigg[\frac{\big(\sum_i\Qc_i\big)^2}{M_f}-\sum_i\frac{\Qc_i^2}{m_i}\bigg]\;}
\label{eq:scalarmemory}
\end{equation}
in terms of the endpoint data alone, with the radiated energy $E_{\rm rad}$ entering through $M_f$. Two cases illustrate its content for a binary with $m_1=m_2=m$. For equal charges, $\Qc_1=\Qc_2=\Qc$, the hair of the remnant is almost exactly the sum of the initial hair, and the memory is controlled by the radiated energy,
\begin{equation}
\Delta\tau_1^{(\ell=0)}=-\frac{i\alpha'\,\Qc^2}{4\,m\,M_f}\,E_{\rm rad}.
\end{equation}
For opposite charges, $\Qc_2=-\Qc_1=-\Qc$, the remnant is uncharged and the scalar hair disappears entirely,
\begin{equation}
\Delta\tau_1^{(\ell=0)}=\frac{i\alpha'\,\Qc^2}{4m}.
\end{equation}

Under the residual $U(1)$ of Eq.~\eqref{eq:U1}, every charge rotates as $\Qc_i\to e^{i\theta}\Qc_i$, so that $\Delta\tau_1\to e^{2i\theta}\Delta\tau_1$, while the metric, the radiated energy, the gravitational waveform, the displacement memory~\eqref{eq:memory} and the spin memory~\eqref{eq:spinmemory} are unchanged. A merger of electrically charged black holes produces pure dilaton memory. Its image under $\theta=\pi/2$ is the merger of magnetically charged black holes, with dilaton memory of the opposite sign, and its image under $\theta=\pi/4$ is a merger of dyonic black holes with pure axion memory. These three processes are indistinguishable through the gravitational memories and differ in the scalar sector, where $\Delta a_1=\re\Delta\tau_1$ and $\Delta\phi_1=-\im\Delta\tau_1/(2\im\tau_0)$ record the duality phase of the charges. Since all the charges in such a process share the same duality phase, the dyonic contribution to the spin memory derived in Sec.~\ref{sec:spin} vanishes, and the spin memories of the three processes coincide as well.

The quantities discussed in this paper have different operational meanings. The displacement memory~\eqref{eq:memory} is the permanent relative displacement of freely falling test masses. The electromagnetic memory~\eqref{eq:EMmemory} is the integrated Lorentz force, i.e.\ the net momentum kick, on charged test particles. The scalar memory $\Delta\tau_1$ is the change of the asymptotic dilaton--axion data; its measurement requires a probe coupled to the dilaton or the axion, and with such a probe it records the duality phase of the sources. The spin memory of Sec.~\ref{sec:spin} is the relative time delay of counter-propagating light rays around a closed contour~\cite{Pasterski:2015tva}. Finally, the $O(\alpha')$ term of Sec.~\ref{sec:alpha} enters the evolution of the spin-two curvature coefficient $\Psi_0^0$, which is the aspect of the spin-two asymptotic charges~\cite{Freidel:2021dfs,Freidel:2021ytz}; it governs the asymptotic curvature and the spin-two charges built from it, and thereby the subleading soft sector of the theory.

%=====================================================================
\section{Spin memory and dyonic sources}
\label{sec:spin}

The spin memory is the magnetic-parity counterpart of the displacement memory~\cite{Pasterski:2015tva,Flanagan:2015pxa,Nichols:2017rqr}. It is measured by the relative delay of light rays sent around a closed contour in opposite directions, and it is controlled by the retarded-time integral of the magnetic-parity potential of the shear in Eq.~\eqref{eq:EBdec},
\begin{equation}
\Sigma(\hat n)\equiv\int_{-\infty}^{\infty}du\;\Ss(u,\hat n).
\end{equation}
The relevant balance law is the magnetic-parity projection of Eq.~\eqref{eq:Nevol}. Applying $\epsilon^{AB}D_A$ removes $D_AM$ and the gradient term, together with the electric-parity part of the linear terms in the shear, and we find on the unit sphere the identities
\begin{equation}
D^Bc_{AB}\big|_{\Ss}=\epsilon_A{}^BD_B(D^2+2)\Ss,\qquad
\epsilon^{AB}D_A\Big(\tfrac14D_CD_BD_Dc^{CD}-\tfrac14D^2D^Cc_{BC}\Big)=\tfrac14D^2D^2(D^2+2)\Ss .
\end{equation}
Integrating in $u$ between nonradiative epochs, we obtain the spin-memory balance law of the heterotic truncation:
\begin{equation}
{\;
\begin{aligned}
\tfrac14D^2D^2(D^2+2)\,\Sigma={}&\Delta\big(\epsilon^{AB}D_AN_B\big)-\int du\;\epsilon^{AB}D_A\mathcal J_B\\
&+\int du\Big[\frac{\epsilon^{AB}\re\big(D_A\partial_u\tau_1\,D_B\bar\tau_1\big)}{2(\im\tau_0)^2}-\frac{\alpha'}{4}\,\frac{\im D^A\big(\bar\Qc\,\psi_A\big)}{\im\tau_0}\Big]
\end{aligned}\;}
\label{eq:spinmemory}
\end{equation}
The gauge term follows from Eq.~\eqref{eq:TuA} and from the self-duality of $\psi_A$, which turns the curl of $\re(\bar\Qc\psi_A)$ into a divergence:
\begin{equation}
\epsilon^{AB}D_A\re\big(\bar\Qc\psi_B\big)=\re D_A\big(\bar\Qc\,\epsilon^{AB}\psi_B\big)=-\im D^A\big(\bar\Qc\psi_A\big).
\label{eq:curlid}
\end{equation}
The operator $\mathcal D\equiv D^2D^2(D^2+2)$ annihilates the $\ell=0,1$ harmonics and has eigenvalue $-\ell^2(\ell+1)^2(\ell-1)(\ell+2)$ on $\ell\ge2$. Equation~\eqref{eq:spinmemory} therefore splits into two independent statements. Its projection $\mathbb P_{\ell=1}$ on the dipole harmonics, for which the left-hand side vanishes, is the balance law of the Bondi angular momentum: the change of the $\ell=1$ magnetic-parity part of $N_A$ equals the angular momentum carried by gravitational, dilaton--axion and electromagnetic radiation. Its projection $\mathbb P_{\ell\ge2}$ determines the spin memory,
\begin{equation}
\Sigma=4\,\mathcal D^{-1}\,\mathbb P_{\ell\ge2}\big[\text{right-hand side of Eq.~\eqref{eq:spinmemory}}\big],
\label{eq:Sigmainv}
\end{equation}
where $\mathcal D^{-1}$ is defined on $\ell\ge2$; the $\ell=0,1$ parts of $\Ss$ do not contribute to the shear. Each matter term is invariant under the full $SL(2,\mathbb R)$ of Eq.~\eqref{eq:SL2data}, and so is the spin memory.

In terms of the real aspects, the curl of the gauge part of Eq.~\eqref{eq:TuA} is, using the Gauss law and the Bianchi identity~\eqref{eq:gauss},
\begin{equation}
8\pi\,\epsilon^{AB}D_AT^{(2)}_{uB}\Big|_{\rm gauge}=\frac{\alpha'}4\,e^{-2\phi_0}\Big(Q\,\partial_uf-f\,\partial_uQ+\epsilon^{AB}D_AQ\,\dot A^{(0)}_B-D^Af\,\dot A^{(0)}_A\Big),
\label{eq:curlreal}
\end{equation}
whose part linear in the conserved monopoles $q_0$, $p_0$ of $Q$ and $f$ is $\tfrac{\alpha'}4e^{-2\phi_0}(q_0\partial_uf-p_0\partial_uQ)=-\tfrac{\alpha'}4\partial_u\langle\Qc_0,\Qc\rangle$. The structure of the gauge source is best displayed with the charge evolution~\eqref{eq:Qevol}, which gives
\begin{equation}
\im D^A\big(\bar\Qc\psi_A\big)=\im\big(\bar\Qc\,\partial_u\Qc\big)+\im\big(\psi^AD_A\bar\Qc\big).
\end{equation}
The complex charge aspect separates into its monopole, $\Qc_0=\frac1{4\pi}\oint d^2\Omega\,\Qc$, which is conserved, and the $\ell\ge1$ part $\Qc'$. The term linear in the monopole is a total derivative in $u$, and its contribution is fixed by the initial and final charge aspects. We accordingly write the spin memory as
\begin{equation}
\Sigma=\Sigma_{\rm ord}+\Sigma_{\rm null},
\end{equation}
where $\Sigma_{\rm ord}$ is the part determined by endpoint data, namely $\Delta(\epsilon^{AB}D_AN_B)$ and the dyonic term, and $\Sigma_{\rm null}$ collects the retarded-time integrals of the gravitational flux $\epsilon^{AB}D_A\mathcal J_B$, of the scalar flux, and of the gauge flux $\im(\bar\Qc'\partial_u\Qc')+\im(\psi^AD_A\bar\Qc')$, which is quadratic in the radiative and anisotropic data. The split is exact and introduces no boundary terms. Since $\partial_u\Qc_0=0$ and $D_A\Qc_0=0$, the monopole enters the gauge source only through $\im(\bar\Qc_0\,\partial_u\Qc')=\partial_u\im(\bar\Qc_0\Qc)$, a total retarded-time derivative whose integral is evaluated at the endpoints $u\to\pm\infty$ and defines the ordinary part. The angular manipulations, the curl identity~\eqref{eq:curlid} and the inversion~\eqref{eq:Sigmainv}, are pointwise identities or act on smooth, globally defined fields on the closed sphere, where $\mathcal D^{-1}$ is unique on $\ell\ge2$. The dyonic part of $\Sigma_{\rm ord}$ is
\begin{equation}
{\;\tfrac14\,\mathcal D\,\Sigma_{\rm ord}\Big|_{\rm dyonic}=-\frac{\alpha'}{4}\,\mathbb P_{\ell\ge2}\big\langle\Qc_0,\Delta\Qc\big\rangle\;}
\label{eq:DSZmemory}
\end{equation}
with the symplectic product~\eqref{eq:DSZ}, while $\mathbb P_{\ell=1}\langle\Qc_0,\Delta\Qc\rangle$ enters the angular-momentum balance. The monopole of $\langle\Qc_0,\Delta\Qc\rangle$ vanishes identically, since $\Qc_0$ is conserved. Equation~\eqref{eq:DSZmemory} is nonzero only when the change of the charge aspect has a nonvanishing Dirac--Schwinger--Zwanziger pairing with the total charge, and it is a genuinely dyonic effect: the static magnetic (electric) Coulomb field of the total charge, combined with the electric (magnetic) radiation that changes the charge aspect, carries angular momentum to null infinity. For charges of a single duality phase, $\Qc_0$ and $\Delta\Qc$ are parallel in the complex plane and this contribution vanishes, while the gravitational, scalar and null electromagnetic contributions to $\Sigma$ remain. {For a monopole embedded in one $E_8$ factor, as in Sec.~\ref{sec:action}, with conserved and angle-independent magnetic aspect $f=p$ and vanishing total electric charge, $\mathsf q_{e,0}=0$, one has $\langle\Qc_0,\Delta\Qc\rangle=p\,\Delta\mathsf q_e$ and
\begin{equation}
\tfrac14\,\mathcal D\,\Sigma_{\rm ord}\Big|_{\rm monopole}=-\frac{\alpha'}{4}\,p\,\mathbb P_{\ell\ge2}\,\Delta\mathsf q_e .
\label{eq:E8monopole}
\end{equation}
The monopole converts the gauge radiation that changes the electric charge aspect into the angular-momentum flux $\tfrac{\alpha'}4e^{-2\phi_0}\dot A^{(0)C}F^{(0)}_{AC}$ of Eq.~\eqref{eq:TuA}, whose curl is the source of Eq.~\eqref{eq:E8monopole}. The electric charge that enters is $\mathsf q_e$ of Eq.~\eqref{eq:qe}, which contains the shift $-\re\tau_0\,p$ induced by the axion coupling $aF\tilde F$ inherited from the Green--Schwarz mechanism, so that the axion vacuum mixes electric and magnetic data without cancelling the monopole contribution.}

For an initial and a final configuration of widely separated bodies with complex charges $\Qc_i$ and velocities $\mathbf v_i$, the charge aspect at null infinity is the boosted Coulomb profile~\cite{Bieri:2013hqa,Strominger:2017zoo}
\begin{equation}
\Qc(\hat n)=\sum_i\Qc_i\,K(\hat n,\mathbf v_i),\qquad K(\hat n,\mathbf v)=\frac{1-v^2}{(1-\hat n\cdot\mathbf v)^2},\qquad \frac1{4\pi}\oint d^2\Omega\,K=1,
\label{eq:LW}
\end{equation}
and Eq.~\eqref{eq:DSZmemory} becomes
\begin{equation}
\tfrac14D^2D^2(D^2+2)\,\Sigma\Big|_{\rm dyonic}=-\frac{\alpha'}{4}\Big[\sum_{a\in{\rm out}}\langle\Qc_{\rm tot},\Qc_a\rangle K(\hat n,\mathbf v_a)-\sum_{b\in{\rm in}}\langle\Qc_{\rm tot},\Qc_b\rangle K(\hat n,\mathbf v_b)\Big],
\label{eq:DSZbodies}
\end{equation}
with $\Qc_{\rm tot}=\Qc_0$ the conserved total charge and the projection $\mathbb P_{\ell\ge2}$ understood. Two consequences are immediate. In the scattering of two bodies, $\langle\Qc_{\rm tot},\Qc_1\rangle=-\langle\Qc_{\rm tot},\Qc_2\rangle=\langle\Qc_2,\Qc_1\rangle$, and the dyonic contribution is the Dirac--Schwinger--Zwanziger pairing of the two bodies times the change of their relative Coulomb profiles. In a merger of initially static bodies into a single remnant, $\langle\Qc_{\rm tot},\Qc_{\rm tot}\rangle=0$ and $\sum_i\langle\Qc_{\rm tot},\Qc_i\rangle=0$, so the ordinary dyonic contribution identified through the monopole--charge-aspect pairing vanishes, whatever the charges and the recoil velocity of the remnant. The spin memory of such a merger is then determined by its gravitational, scalar and null electromagnetic contributions.

The normalization of Eq.~\eqref{eq:DSZmemory} is fixed by that of the charges in the action: in terms of the charge aspects of Eq.~\eqref{eq:qe}, $\langle\Qc_1,\Qc_2\rangle=f_1\mathsf q_{e2}-\mathsf q_{e1}f_2$, so that $\alpha'\langle\Qc_1,\Qc_2\rangle/8$ is the Dirac--Schwinger--Zwanziger product of the canonically normalized charges. Its dipole part provides an independent check of this normalization, as we now show.

The dyonic spin memory is generated in the scattering of mutually nonlocal objects. Consider two of the black holes of Sec.~\ref{sec:memory} with charges $\Qc_1$, $\Qc_2$ and $\langle\Qc_1,\Qc_2\rangle\neq0$, e.g.\ an electrically and a magnetically charged black hole, that scatter with incoming and outgoing velocities $\mathbf v_i^{\mp}$. Equation~\eqref{eq:DSZbodies} gives
\begin{equation}
\tfrac14D^2D^2(D^2+2)\,\Sigma\Big|_{\rm dyonic}=-\frac{\alpha'}{4}\,\langle\Qc_1,\Qc_2\rangle\Big[\Delta K(\hat n,\mathbf v_2)-\Delta K(\hat n,\mathbf v_1)\Big],\qquad
\Delta K(\hat n,\mathbf v_i)=K(\hat n,\mathbf v^+_i)-K(\hat n,\mathbf v^-_i).
\end{equation}
At low velocities, $K=(1-v^2)\sum_{k\ge0}(k+1)(\hat n\cdot\mathbf v)^k=1+2\,\hat n\cdot\mathbf v+3(\hat n\cdot\mathbf v)^2-v^2+O(v^3)$. Its dipole part is $2\,\hat n\cdot\mathbf v$ up to corrections of order $v^3$, and its quadrupole part is $3\,\hat n_{\langle i}\hat n_{j\rangle}v^iv^j$ up to corrections of order $v^4$, where $\langle ij\rangle$ denotes the symmetric trace-free part.

The dipole part enters the angular-momentum balance. The angular momentum about the unit vector $\mathbf e$ carried to null infinity is the flux of $T^r{}_A$ through a large sphere, $J^{\rm out}_{\mathbf e}=-\int du\oint d^2\Omega\;Y^AT^{(2)}_{uA}$ with $Y^A=\epsilon^{AB}D_B(\hat n\cdot\mathbf e)$ the rotation about $\mathbf e$ and $g^{ru}=-1$ at leading order. Using $\oint Y^AV_A=\oint(\hat n\cdot\mathbf e)\,\epsilon^{AB}D_AV_B$ and Eq.~\eqref{eq:curlid}, the dyonic part is
\begin{equation}
\mathbf J^{\rm out}_{\rm dyonic}=\frac{\alpha'}{32\pi}\oint d^2\Omega\;\hat n\,\big\langle\Qc_0,\Delta\Qc\big\rangle=\frac{\alpha'}{12}\,\langle\Qc_1,\Qc_2\rangle\,\Delta\big(\mathbf v_2-\mathbf v_1\big)+O(v^3),
\label{eq:DeltaJ}
\end{equation}
directed along the change of the relative velocity. Equation~\eqref{eq:DeltaJ} follows from the balance law, and its coefficient and sign can be tested independently in flat space. Consider a charge with Coulomb field $q\,\hat{\mathbf r}/r^2$, scattered with a change of velocity $\Delta\mathbf v$ off a static monopole with field $g\,\hat{\mathbf r}/r^2$, in a theory whose stress tensor is $\kappa/(4\pi)$ times the Maxwell one, with $\kappa=\alpha'\im\tau_0/8$ in the normalization of Eq.~\eqref{eq:action}. The angular-momentum flux of the cross term between the monopole field and the Larmor radiation of the charge, $\frac{\kappa g}{4\pi}\oint d^2\Omega\int dt\,r\mathbf E_{\rm rad}$ with $r\mathbf E_{\rm rad}=-q\,\mathbf a_\perp$, gives $-\tfrac23\kappa\,q\,g\,\Delta\mathbf v$. With $A_t=-\Phi$, the aspects are $Q=-q$ and $f=g$, so that $\langle\Qc_{\rm mono},\Qc_{\rm charge}\rangle=-\im\tau_0\,g\,q$, and Eq.~\eqref{eq:DeltaJ} reproduces this value in magnitude and sign. The dyonic flux is the radiative counterpart of the angular momentum stored in the Coulomb field of a charge--monopole pair.

The quadrupole part, on which $\tfrac14\mathcal D=-36$, gives, with $\langle\Qc_{\rm tot},\Qc_2\rangle=-\langle\Qc_{\rm tot},\Qc_1\rangle=\langle\Qc_1,\Qc_2\rangle$ in Eq.~\eqref{eq:DSZbodies}, the leading dyonic contribution to the spin memory,
\begin{equation}
\Sigma^{(\ell=2)}_{\rm ord}\Big|_{\rm dyonic}=\frac{\alpha'}{48}\,\langle\Qc_1,\Qc_2\rangle\;\hat n_{\langle i}\hat n_{j\rangle}\,\Delta\big(v_2^iv_2^j-v_1^iv_1^j\big)+O(v^4).
\label{eq:Sigmal2}
\end{equation}
It is fixed by the Dirac--Schwinger--Zwanziger pairing of the scattered black holes and by the change of their velocity quadrupole, and it is invariant under the full duality group. The gravitational and null contributions to the spin memory of the same process add to it.

%=====================================================================
\section{Order \texorpdfstring{$\alpha'$}{alpha'}: the first curvature correction at null infinity}
\label{sec:alpha}

We now turn to the curvature couplings~\eqref{eq:Salpha}, working within the four-derivative effective action~\eqref{eq:Seff} and to first order in $\alpha'$. On the radiative background of Sec.~\ref{sec:asymptotics}, the tensors $H_{\mu\nu}$ and $C_{\mu\nu}$ are linear in the second derivatives of the scalars and in the curvature. The dominant component of the Hessian of the scalars, $\nabla_r\nabla_r$ in coordinate form, is contracted with the principal null direction of the radiative Weyl tensor, which annihilates $\Psi_4$, $\Psi_3$ and their duals. As a result both tensors decay faster than the matter stress tensor:
\begin{gather}
H_{uu},\,C_{uu}=O(r^{-4}),\qquad H_{uA},\,C_{uA}=O(r^{-3}),\qquad H_{rA},\,C_{rA}=O(r^{-4}),\notag\\
H_{ur},\,C_{ur}=O(r^{-5}),\qquad H_{rr}=O(r^{-6}),\qquad C_{rr}=O(r^{-5}),\qquad H^{\rm TF}_{AB},\,C^{\rm TF}_{AB}=O(r^{-2}).
\label{eq:HCfalloff}
\end{gather}
Each entry lies between one and three powers of $1/r$ below the bounds of Ref.~\cite{Maibach:2026afb}, so that the curvature couplings preserve the smooth expansion and the conditions for the absence of logarithms. The leading coefficients, together with an order-by-order account of the contractions that produce them, are given in Appendix~\ref{app:curvature}. Since $T_{uu}$ and $T_{uA}$ enter the balance laws at $O(r^{-2})$, we obtain the following statement: within the four-derivative effective theory and to first order in $\alpha'$, the explicit Gauss--Bonnet and Pontryagin contributions do not modify the Bondi mass loss~\eqref{eq:massloss}, the angular-momentum balance~\eqref{eq:Nevol}, the charge evolution~\eqref{eq:Qevol}, or the displacement and spin-memory formulas~\eqref{eq:memory} and~\eqref{eq:spinmemory}, for the smooth asymptotic expansion~\eqref{eq:metricfalloff}. The data entering these laws, such as the news, $\tau_1$ and $\psi_A$ of a given source, receive their own $O(\alpha')$ corrections, which act through the same balance laws. The densities $\GB$ and $\Po$ are quadratic in the Weyl scalars and in the Ricci tensor and decay as $r^{-6}$, so they first enter the scalar hierarchy in the evolution of $\tau_5$.

The first dynamical equation of the hierarchy that contains an explicit $\alpha'$ term is the trace-free angular equation at $O(r^{-2})$, which determines the retarded-time derivative of the trace-free $O(r^{-3})$ coefficient $e_{AB}$ of $h_{AB}$. The leading trace-free angular components of the curvature tensors are
\begin{equation}
H^{\rm TF}_{AB}\big[e^{-2\phi}\big]=\frac{2e^{-2\phi_0}\,\phi_1\,\partial_uN_{AB}}{r^2}+O(r^{-3}),\qquad
C^{\rm TF}_{AB}[a]=\frac{a_1\,{}^\star\partial_uN_{AB}}{r^2}+O(r^{-3}),
\label{eq:HCTF}
\end{equation}
where ${}^\star X_{AB}\equiv\epsilon_A{}^CX_{CB}$ is the magnetic-parity rotation of a trace-free tensor on the sphere. In $H_{AB}=P_{AcBd}\nabla^c\nabla^df$ the only contraction that reaches $O(r^{-2})$ is $P_{AuBu}f^{uu}$, with
\begin{equation}
f^{uu}=(g^{ur})^2\nabla_r\nabla_rf=-\frac{4e^{-2\phi_0}\phi_1}{r^3}+O(r^{-4}),\qquad P_{AuBu}=R_{AuBu}+O(r^0)=-\frac r2\,\partial_uN_{AB}+O(r^0);
\end{equation}
every other contraction is $O(r^{-3})$ or smaller. In $C_{AB}$ the corresponding contribution comes from the term $\nabla_\sigma\nabla_\tau\vartheta\,{}^\star R^{\tau(\mu\nu)\sigma}$, while the term with $\nabla R_{\mu\nu}$ is suppressed by the falloff of the Ricci tensor. Both coefficients are products of the scalar charge aspect and the retarded-time derivative of the news, i.e.\ of the scalar data and of $\Psi_4$. The subleading components follow the same pattern, e.g.
\begin{equation}
H^{(3)}_{uA}\big[e^{-2\phi}\big]=-2e^{-2\phi_0}\big[\partial_uN_{AB}D^B\phi_1-\dot\phi_1D^BN_{AB}\big],\qquad
C^{(3)}_{uA}[a]=-\big[{}^\star\partial_uN_{AB}D^Ba_1-\dot a_1\,\epsilon_A{}^CD^BN_{BC}\big].
\end{equation}

The complete trace-free angular equation at $O(r^{-2})$, including the gravitational, scalar and gauge terms, reads
\begin{equation}
{\;
\begin{aligned}
\partial_ue_{AB}={}&\tfrac13D_{\langle A}N_{B\rangle}+\tfrac12Mc_{AB}+\tfrac14c_{AB}\,c_{CD}N^{CD}-\tfrac18c_{AB}D^CD^Dc_{CD}+\tfrac13D_{\langle A}\big(c_{B\rangle C}D_Dc^{CD}\big)\\
&+\tfrac1{16}\big(c^{CD}D_{\langle A}D_{B\rangle}c_{CD}-D_{\langle A}c_{CD}\,D_{B\rangle}c^{CD}\big)\\
&+\frac{\re\big(\tau_1D_{\langle A}D_{B\rangle}\bar\tau_1-D_{\langle A}\tau_1D_{B\rangle}\bar\tau_1\big)}{8(\im\tau_0)^2}
-\frac{\alpha'}{4}e^{-2\phi_0}\,\dot A^{(0)}_{\langle A}A^{(1)}_{B\rangle}\\
&+\frac{\alpha'}{2}\,e^{-2\phi_0}\,\phi_1\,\partial_uN_{AB}-\frac{\alpha'}{4}\,a_1\,{}^\star\partial_uN_{AB},
\end{aligned}\;}
\label{eq:efull}
\end{equation}
where $\langle AB\rangle$ denotes the symmetric trace-free part on the unit sphere, $A^{(1)}_A$ is the $O(r^{-1})$ coefficient of the angular gauge potential in radial gauge, and $N_A$ is the angular-momentum aspect of Eq.~\eqref{eq:Nevol}. The first three lines are the two-derivative equation, in which the $\alpha'$ of the gauge term is the normalization of Sec.~\ref{sec:action}; the scalar term is duality invariant. The last line is the contribution of the curvature couplings,
\begin{equation}
\partial_ue_{AB}\Big|_{\alpha'}=\frac{\alpha'}{2}\,e^{-2\phi_0}\,\phi_1\,\partial_uN_{AB}-\frac{\alpha'}{4}\,a_1\,{}^\star\partial_uN_{AB}.
\label{eq:eevol}
\end{equation}
Its two terms arise from $-\tfrac{\alpha'}{2}H_{AB}$ and $+\tfrac{\alpha'}{2}C_{AB}$ in Eq.~\eqref{eq:einstein} with the leading coefficients~\eqref{eq:HCTF}, and their relative weight is fixed by the action~\eqref{eq:Salpha}, i.e.\ by the combination ${\rm Tr}F^2-{\rm tr}R^2$ and the Lorentz Chern--Simons term. No other term in Eq.~\eqref{eq:efull} depends on the couplings: the matter terms and the lower-order metric coefficients are fixed at orders where $H_{\mu\nu}$ and $C_{\mu\nu}$ do not contribute, by Eq.~\eqref{eq:HCfalloff}, and the operators in $S_{\rm other}$ do not reach this order (Appendix~\ref{app:operators}).

We translate Eq.~\eqref{eq:eevol} into the Newman--Penrose formalism~\cite{Newman:1961qr,Newman:1962cia} with the tetrad
\begin{equation}
l=e^{-2\beta}\partial_r,\qquad n=\partial_u-\frac V{2r}\partial_r+U^A\partial_A,\qquad m=\frac1rM^A\partial_A,
\label{eq:tetrad}
\end{equation}
where $M^A$ is a complex null dyad of $h_{AB}$, $h_{AB}M^AM^B=0$, $h_{AB}M^A\bar M^B=1$, which reduces at leading order to $\hat m=\tfrac1{\sqrt2}(\partial_\theta+\tfrac{i}{\sin\theta}\partial_\varphi)$. With the metric~\eqref{eq:bondi} one checks that
\begin{equation}
l\cdot l=n\cdot n=m\cdot m=l\cdot m=n\cdot m=0,\qquad l\cdot n=-1,\qquad m\cdot\bar m=1
\end{equation}
hold exactly. For the orientation $\epsilon_{\theta\varphi}=\sin\theta$, the dyad satisfies $\epsilon_A{}^C\hat m^A=-i\hat m^C$, and hence ${}^\star X_{AB}\hat m^A\hat m^B=-iX_{AB}\hat m^A\hat m^B$ for any trace-free $X_{AB}$. Since $g(l,l)=g(l,m)=g(m,m)=0$, $\Psi_0=C_{abcd}l^am^bl^cm^d$ coincides with the corresponding component of the Riemann tensor, $\Psi_0=e^{-4\beta}R_{rArB}m^Am^B$, and the leading coefficients are
\begin{equation}
\Psi_0^0=-3\,e_{AB}\,\hat m^A\hat m^B-\frac{|\tau_1|^2}{8(\im\tau_0)^2}\,c_{AB}\,\hat m^A\hat m^B+\big(\text{terms cubic in }c_{AB}\big),
\qquad
\Psi_4^0=-\tfrac12\,\partial_uN_{AB}\,\hat{\bar m}^A\hat{\bar m}^B .
\label{eq:Psi0lin}
\end{equation}
The nonlinear terms in $\Psi_0^0$ involve only the free data $c_{AB}$ and $\tau_1$, whose evolution contains no explicit $\alpha'$ term, so that
\begin{equation}
\partial_u\Psi_0^0\Big|_{\alpha'}=-3\,\partial_ue_{AB}\Big|_{\alpha'}\hat m^A\hat m^B
=-\frac{3\alpha'}{4}\big(2e^{-2\phi_0}\phi_1+ia_1\big)\,\partial_uN_{AB}\hat m^A\hat m^B .
\end{equation}
Since $N_{AB}$ is real, $\bar\Psi_4^0=-\tfrac12\partial_uN_{AB}\hat m^A\hat m^B$, and $2e^{-2\phi_0}\phi_1+ia_1=i(a_1-2ie^{-2\phi_0}\phi_1)=i\tau_1$. Therefore
\begin{equation}
{\;\partial_u\Psi_0^0\Big|_{\alpha'}=\frac{3i\alpha'}{2}\,\tau_1\,\bar\Psi_4^0\;}
\label{eq:Psi0alpha}
\end{equation}
The dilaton and axion contributions combine into the complex scalar charge aspect $\tau_1$ as a direct consequence of the holomorphic structure $\im[\tau(\GB+i\Po)]$ of Eq.~\eqref{eq:Salpha}. The rest of $\partial_u\Psi_0^0$ is the asymptotic Bianchi identity $\eth\Psi_1^0+3\sigma^0\Psi_2^0$ of Newman and Unti~\cite{Newman:1962cia} together with the minimally coupled matter terms of Eq.~\eqref{eq:efull}; the evolution equations of $\Psi_2^0$ and $\Psi_1^0$, equivalent to Eqs.~\eqref{eq:massloss} and~\eqref{eq:Nevol}, contain no explicit $\alpha'$ term.

Equation~\eqref{eq:Psi0alpha} has the following properties. (i) It is a first-order effect in the expansion parameter $\alpha'/M^2$. For generic radiative data $\tau_1$ is of order one, and for the black holes of Sec.~\ref{sec:memory} with fixed physical charges it is of order one as well, Eq.~\eqref{eq:tau1can}; the factor $\alpha'$ in Eq.~\eqref{eq:tau1BH} only reflects the normalization of the gauge fields. This distinguishes the heterotic truncation from scalar--Gauss--Bonnet or dynamical Chern--Simons models without a gauge sector, where the scalar charge of a black hole is itself induced by the curvature coupling and the analogous term is of second order. (ii) It is invariant under the axion shifts $\tau\to\tau+b$, which leave $\tau_1$ unchanged, as required by the topological nature of $\tau_0(\GB+i\Po)$. (iii) Within the four-derivative truncation and with the two-derivative duality transformations~\eqref{eq:U1}, it is not invariant under the residual rotation, under which $\tau_1\to e^{2i\theta}\tau_1$ while $\Psi_0$ and $\Psi_4$ are unchanged; the S-dual processes of Sec.~\ref{sec:memory}, which share their two-derivative gravitational radiation, enter this equation with different phases of $\tau_1$ relative to the news. (iv) It is covariant under the spin rotations $\hat m\to e^{i\lambda}\hat m$ of the dyad, under which both sides carry spin weight two, and under BMS transformations, under which $\tau_1$ and $\Psi_4^0$ transform as fields at null infinity of definite spin weight.

The coefficient $\Psi_0^0$ is the aspect of the spin-two asymptotic charges~\cite{Freidel:2021dfs,Freidel:2021ytz}, and Eq.~\eqref{eq:Psi0alpha} is the first correction from the curvature couplings to its evolution equation. Integrated between nonradiative epochs, where the news vanishes, it gives the net change of the spin-two aspect
\begin{equation}
\Delta\Psi_0^0\Big|_{\alpha'}=\frac{3i\alpha'}{2}\int du\;\tau_1\bar\Psi_4^0=\frac{3i\alpha'}{4}\int du\;\partial_u\tau_1\,N_{AB}\,\hat m^A\hat m^B,
\label{eq:DeltaPsi0}
\end{equation}
a cross term between the scalar news and the gravitational news. It is generated in processes in which scalar and gravitational radiation are emitted together, such as the black-hole processes of Secs.~\ref{sec:memory} and~\ref{sec:spin}.

%=====================================================================
\section{Conclusions}
\label{sec:conclusions}

We have determined the asymptotic structure at null infinity of the bosonic sector of the heterotic string compactified on $T^6$, in the consistent truncation to the metric, the axion--dilaton and the gauge fields from ten dimensions, through first order in $\alpha'$. The dilaton, the axion and the gauge fields each carry free radiative data and contribute at leading order to the Bondi mass loss, to the angular-momentum balance and to the displacement and spin memories. The organization of these contributions is dictated by the duality symmetry of the four-dimensional theory: the scalar flux is the metric of the dilaton--axion moduli space evaluated on the scalar news, the gauge flux is the norm of a self-dual radiation field, and all balance laws and memory formulas of the two-derivative theory are covariant under S-duality acting together with the vacuum value of the axion--dilaton. At fixed vacuum the residual rotation acts with charges zero, one and two on gravitational, electromagnetic and dilaton--axion radiation, which identifies precisely what distinguishes S-dual processes at null infinity: their scalar data, and nothing in their gravitational radiation.

The spin memory reveals a genuinely dyonic effect. Its ordinary gauge-field part is governed by the Dirac--Schwinger--Zwanziger product of the total charge with the change of the charge aspect. It is generated in the scattering of mutually nonlocal objects, with an amplitude set by their duality-invariant symplectic product and by the change of their velocity quadrupole, and its dipole part agrees in magnitude and sign with the angular momentum radiated by a charge scattered off a magnetic monopole. {For a monopole embedded in one $E_8$ factor, whose magnetic charge is a topological integer of the reduction $E_8\to(E_7\times U(1))/\mathbb Z_2$ on the celestial sphere, it is the product of this charge with the change of the electric charge aspect.} For mergers of initially static black holes this ordinary dyonic part vanishes, while the gravitational, scalar and null electromagnetic contributions remain. The effect ties the infrared structure of the heterotic string to the quantization of electric and magnetic charges and to the angular momentum stored in the Coulomb fields of dyons.

For processes connecting stationary black holes of the theory, including the rotating ones, the scalar memory is fixed by the charges and masses of the initial and final states, and S-dual processes share their gravitational waveform and memories while their scalar memories carry the square of the duality phase. Expressed through the physical charges, the scalar charge of these black holes is of order one in the inverse string tension and comparable to the mass near extremality.

Within the four-derivative effective action, the Gauss--Bonnet and Pontryagin couplings required by the Green--Schwarz mechanism leave the mass-loss and angular-momentum balance laws without explicit corrections at first order in $\alpha'$, and first enter the evolution of the spin-two curvature coefficient. The correction is first order in $\alpha'/M^2$, combines dilaton and axion into the complex scalar charge aspect, and is not invariant under the residual duality rotation of the two-derivative theory. It singles out the level of the asymptotic hierarchy at which string corrections appear explicitly at null infinity, and links them to the spin-two charges that organize the subleading soft sector.

These results open several directions. The endpoint relations for black-hole processes call for a dynamical counterpart: the post-Minkowskian or numerical evolution of black-hole binaries in the heterotic truncation would supply the waveforms that feed the null contributions to the memories, and would turn the scalar and dyonic spin memories into complete predictions for scattering and merger events. The construction of the associated charges through the covariant phase-space formalism of Wald and Zoupas~\cite{Wald:1999wa}, combined with a duality-invariant treatment of the magnetic sector~\cite{Mitsios:2021zrn,Hosseinzadeh:2018dkh,Freidel:2018fsk}, would organize the electric and magnetic large gauge transformations and the dual symmetries of the axion~\cite{Campiglia:2018see,Francia:2018jtb,Seraj:2021rxd} into multiplets of the residual rotation and give a charge-algebra interpretation of the dyonic spin memory. The same framework, applied to the spin-two charges, would turn the correction to the evolution of $\Psi_0$ into a correction to an operationally defined observable, and the next order in $\alpha'$ would determine how the full string effective action, beyond the four-derivative truncation, realizes the duality group at null infinity. The inclusion of the moduli and of the Kaluza--Klein and winding vectors would extend the duality group to $SL(2,\mathbb R)\times O(6,22)$ and connect the present analysis with Ref.~\cite{Ferko:2021bym}. Finally, the scalar, two-form and spin memories obtained here are the classical counterparts of the soft theorems for the dilaton and the Kalb--Ramond field in string amplitudes~\cite{DiVecchia:2015oba}, and the curvature correction to the evolution of $\Psi_0$ suggests a corresponding string correction to the sub-subleading soft graviton theorem.

\begin{acknowledgments}
Maickol Muñoz acknowledges the Agencia Nacional de Investigación y Desarrollo (ANID) of Chile through the Programa Ciencia e Innovación para el 2030 at Universidad de La Frontera, for funding the research stay at Universidad de La Rioja. Maickol Muñoz also thanks María del Pilar García del Moral for her warm hospitality, valuable mentorship, and insightful scientific discussions.
\end{acknowledgments}
\section*{Tool Use Declaration}
The authors declare that no generative AI tools or sophisticated automated software were used in the writing, editing, or preparation of this manuscript.

%=====================================================================
\appendix
\section{Consistency of the truncation}
\label{app:truncation}

In the string frame, the four-dimensional action of the heterotic string on $T^6$ is~\cite{Maharana:1992my,Sen:1994fa}
\begin{equation}
S=\frac{1}{32\pi}\int d^4x\sqrt{-G}\,e^{-\Phi}\Big[R_G+G^{\mu\nu}\partial_\mu\Phi\partial_\nu\Phi-\tfrac1{12}H_{\mu\nu\rho}H^{\mu\nu\rho}+\tfrac18{\rm Tr}\big(\partial_\mu M\,L\,\partial^\mu M\,L\big)-\tfrac14F^a_{\mu\nu}\big(LML\big)_{ab}F^{b\mu\nu}\Big],
\label{eq:MS}
\end{equation}
with $H_{\mu\nu\rho}=\partial_\mu B_{\nu\rho}+\tfrac12A^a_\mu L_{ab}F^b_{\nu\rho}+\text{cyclic}$, $a=1,\dots,28$, and $M$ the symmetric $O(6,22)$ matrix built from the internal metric $G_{mn}$, the internal two-form $B_{mn}$ and the Wilson lines $\mathsf a^I_m$. The vectors split into the twelve Kaluza--Klein and winding vectors, $a=m$, and the sixteen vectors from the ten-dimensional gauge fields, $a=I$, and $L={\rm diag}(\sigma_1\otimes\mathbb 1_6,-\mathbb 1_{16})$ is block diagonal in this split. Two properties of $M$ are used: its $16\times16$ block is $\mathbb 1+\mathsf aG^{-1}\mathsf a^T$, quadratic in the Wilson lines and independent of $G_{mn}$ and $B_{mn}$ at $\mathsf a=0$, and its blocks mixing the two sets of vectors are linear in $\mathsf a$. The truncation sets $G_{mn}$ and $B_{mn}$ to constants, $\mathsf a^I_m=0$, and $A^m_\mu=0$, and keeps $G_{\mu\nu}$, $\Phi$, $B_{\mu\nu}$ and $A^I_\mu$.

\emph{Moduli.} The kinetic term of $M$ is quadratic in $\partial M$, so its contribution to the field equations of $G_{mn}$, $B_{mn}$ and $\mathsf a^I_m$ vanishes for constant $M$. The only other dependence on the moduli is through $(LML)_{ab}$; since $F^m=0$, the source term is $-\tfrac14e^{-\Phi}F^I_{\mu\nu}F^{J\mu\nu}\,\partial(LML)_{IJ}/\partial\chi$ with $\chi\in\{G_{mn},B_{mn},\mathsf a^I_m\}$, and it vanishes at $\mathsf a=0$ because $(LML)_{IJ}=\mathbb 1+\mathsf aG^{-1}\mathsf a^T$ is quadratic in $\mathsf a$ and independent of $G$ and $B$. The Kalb--Ramond term does not depend on $M$.

\emph{Kaluza--Klein and winding vectors.} Their field equations are $\nabla_\mu\big(e^{-\Phi}(LML)_{mb}F^{b\mu\nu}\big)$ plus terms proportional to $e^{-\Phi}H^{\mu\nu\rho}L_{mb}F^b_{\mu\rho}$. With $F^m=0$ only $b=I$ contributes, and $(LML)_{mI}$ vanishes at $\mathsf a=0$ while $L_{mI}=0$ identically. Hence $\delta S/\delta A^m_\mu=0$.

Together with the dilaton, Kalb--Ramond and metric equations, which are those of the truncated action, this shows $\delta S/\delta\Phi_{\rm discarded}|_{\Phi_{\rm discarded}=0}=0$ for arbitrary configurations of the retained fields. Passing to the Einstein frame $g_{\mu\nu}=e^{-\Phi}G_{\mu\nu}$ with $\Phi=2\phi$ and dualizing $B_{\mu\nu}$ gives Eq.~\eqref{eq:action}, up to the normalization of $A^I_\mu$.

\emph{Non-abelian gauge fields.} In ten dimensions the truncation keeps a block-diagonal metric with flat internal torus, $\Phi$, $B_{\mu\nu}$ and gauge fields $A_\mu$ independent of the internal coordinates $y^m$, with $A_m=0$. The internal and mixed components of the string-frame equations $R_{MN}+2\nabla_M\nabla_N\Phi-\tfrac14H_{MPQ}H_N{}^{PQ}-\tfrac{\alpha'}{4}{\rm Tr}F_{MP}F_N{}^P=0$ are then identically satisfied, since $R_{mN}=0$, $\nabla_m\nabla_N\Phi=0$, $H_{mPQ}=0$ and $F_{mP}=0$; the trace of the gauge stress tensor enters the dilaton equation only, which is retained. The internal components of the Yang--Mills and Kalb--Ramond equations vanish for the same reason. The truncation is therefore consistent for an arbitrary four-dimensional configuration of the full ten-dimensional gauge group.

\emph{Order $\alpha'$.} The four-derivative terms of the ten-dimensional action~\cite{Bergshoeff:1989de,Metsaev:1987zx} are scalars built from the ten-dimensional curvature with torsion, the Kalb--Ramond field strength including its Chern--Simons terms, and the dilaton. On the truncated configurations, the only background tensors with internal indices are $G_{mn}$ and its inverse. A term in the reduced action that is linear in a discarded fluctuation, $\delta G_{\mu m}$, $\delta B_{\mu m}$, $\delta A_m$, or the trace-free parts of $\delta G_{mn}$ and $\delta B_{mn}$, would require a contraction of an odd number of internal indices, or of a trace-free internal tensor, with these background tensors, and therefore vanishes. The trace of $\delta G_{mn}$ enters the reduced four-derivative terms only through $\sqrt{\det G_{mn}}\,e^{-2\Phi_{10}}=e^{-\Phi}$, i.e.\ through the retained four-dimensional dilaton. The truncation is therefore consistent at $O(\alpha')$, and the reduced four-derivative action takes the form~\eqref{eq:Seff}.

\section{Four-derivative operators and field redefinitions}
\label{app:operators}

\emph{Scheme.} We use the scheme of Bergshoeff and de Roo~\cite{Bergshoeff:1989de}, consistent with the two-loop sigma-model results of Ref.~\cite{Metsaev:1987zx}, in which the ten-dimensional string-frame action through $O(\alpha')$ is
\begin{equation}
\begin{aligned}
S_{10}=\frac1{2\kappa_{10}^2}\int d^{10}x\sqrt{-G}\,e^{-2\Phi_{10}}\Big[&R+4(\partial\Phi_{10})^2-\tfrac1{12}H_{MNP}H^{MNP}\\
&-\tfrac{\alpha'}8\Big({\rm Tr}F_{MN}F^{MN}-R_{MNAB}(\Omega_-)R^{MNAB}(\Omega_-)\Big)\Big],
\end{aligned}
\label{eq:S10}
\end{equation}
with $H=dB+\tfrac{\alpha'}4\big(\omega_L(\Omega_-)-\omega_{YM}\big)$ and the torsionful connection $\Omega_{\pm}=\omega\pm\tfrac12H$. On the truncated configurations of Appendix~\ref{app:truncation} every term reduces to its four-dimensional counterpart, with $\sqrt{-G}\,e^{-2\Phi_{10}}\to\sqrt{-G_4}\,e^{-\Phi}$. The complete set of four-derivative operators of the reduced theory is therefore generated by: (i) ${\rm Tr}F^2$, which is the gauge kinetic term of Eq.~\eqref{eq:action}; (ii) the term $\tfrac{\alpha'}4\omega_{YM}$ in $H$, which after dualization gives the coupling $aF\tilde F$ of Eq.~\eqref{eq:action}; (iii) the term $\tfrac{\alpha'}4\omega_L$ in $H$, which after dualization gives $a\Po$ in Eq.~\eqref{eq:Salpha} together with the difference $\omega_L(\Omega_-)-\omega_L(\omega)$; (iv) $R_{MNAB}(\Omega_-)^2$; (v) the change from the string to the Einstein frame. Terms quadratic in the Chern--Simons forms are $O(\alpha'^2)$.

\emph{Decomposition.} With $R(\Omega_-)=R(\omega)+\mathcal D H+H\wedge H$ schematically,
\begin{equation}
R_{\mu\nu\rho\sigma}(\Omega_-)^2=R_{\mu\nu\rho\sigma}R^{\mu\nu\rho\sigma}+c_1R^{\mu\nu\rho\sigma}\nabla_\rho H_{\sigma\mu\nu}+c_2(\nabla H)^2+c_3R^{\mu\nu\rho\sigma}H_{\mu\nu\lambda}H_{\rho\sigma}{}^\lambda+c_4H^4 ,
\end{equation}
with numerical $c_i$. The cross term is a Ricci-type operator: $R^{\mu\nu\rho\sigma}H_{\sigma\mu\nu}=0$ by the algebraic Bianchi identity, so that after integration by parts it equals $-(\nabla_\rho R^{\mu\nu\rho\sigma})H_{\sigma\mu\nu}$, and $\nabla_\rho R^{\mu\nu\rho\sigma}$ is a combination of derivatives of the Ricci tensor. In four dimensions $H^{\mu\nu\lambda}=e^{4\phi}\epsilon^{\mu\nu\lambda\kappa}\partial_\kappa a$ on shell, and the contraction of two Levi-Civita tensors turns $R^{\mu\nu\rho\sigma}H_{\mu\nu\lambda}H_{\rho\sigma}{}^\lambda$ into a multiple of $G_{\kappa\kappa'}\partial^\kappa a\,\partial^{\kappa'}a$. The term $(\nabla H)^2$ reduces to $(\nabla_\mu\nabla_\nu a)^2$ and Ricci-type terms, and $(\nabla_\mu\nabla_\nu a)^2=(\Box a)^2-R_{\mu\nu}\nabla^\mu a\nabla^\nu a$ up to a total derivative. The term $H^4$ is quartic in $\partial a$. Finally $R_{\mu\nu\rho\sigma}R^{\mu\nu\rho\sigma}=\GB+4R_{\mu\nu}R^{\mu\nu}-R^2$. In the reduction of $a\,{\rm tr}R(\Omega_-)\wedge R(\Omega_-)$, the difference from $a\Po$ is $a\,dY$ with $Y$ built from $H$ and the curvature, i.e.\ $-da\wedge Y$. The frame change produces terms with first and second derivatives of $\phi$ multiplying the curvature or other derivatives of $\phi$. All these operators are classified by derivative counting, with $H$ and $\partial\tau$ counting as one derivative. An operator with one curvature tensor and two first derivatives of the scalars has no further derivatives; its parity-even contractions involve the Ricci tensor only, since $R_{\mu\nu\rho\sigma}$ must be contracted on one index pair with the metric, and its parity-odd contractions, after the four-dimensional duality $H\propto{}^\star da$ is used, reduce to the same form or vanish by the algebraic Bianchi identity. Operators with two first and one second derivative of the scalars, such as $(\partial\phi)^2\Box\phi$ or $\partial^\mu\phi\partial^\nu\phi\nabla_\mu\nabla_\nu\phi=-\tfrac12\Box\phi(\partial\phi)^2$ up to a total derivative, are proportional to the field equations up to quartic terms, and squares of second derivatives reduce to $(\Box\phi)^2$, Ricci-type and quartic operators as above.

\emph{Field redefinitions.} Operators proportional to the Ricci tensor or to the lowest-order field equations are removed by $g_{\mu\nu}\to g_{\mu\nu}+\alpha'\delta g_{\mu\nu}$ and $\tau\to\tau+\alpha'\delta\tau$. On shell, $\delta g_{\mu\nu}$ is a combination of $\partial_{(\mu}\tau\partial_{\nu)}\bar\tau/(\im\tau)^2$, $F_{\mu\rho}F_\nu{}^\rho$ and $g_{\mu\nu}$ times their traces, and $\delta\tau$ is quadratic in first derivatives. After this step $S_{\rm other}$ consists only of operators at least quartic in the first derivatives of the scalars, including the terms quartic in $H$.

\emph{Component estimates.} On the asymptotic solution $\partial_u\tau,\partial_A\tau=O(r^{-1})$, $\partial_r\tau=O(r^{-2})$, $\nabla\tau\cdot\nabla\bar\tau=O(r^{-3})$, $F_{uA}=O(1)$, $F_{ur},F_{AB}=O(r^{-2})$ and $O(1)$ respectively in coordinate components, and the largest curvature component is $R_{uAuB}=O(r)$. The retained couplings $e^{-2\phi}\GB$ and $a\Po$ contribute to the trace-free angular equation at $O(r^{-2})$, Eq.~\eqref{eq:HCTF}, and to $T_{uu}$ and $T_{uA}$ only at $O(r^{-4})$ and $O(r^{-3})$; the operators quartic in first derivatives contribute to $T^{\rm TF}_{AB}$, $T_{uu}$ and $T_{uA}$ at $O(r^{-5})$. In the trace-free angular equation the coefficient $e_{AB}$ evolves at $O(r^{-2})$ in coordinate components, and the balance laws involve $T_{uu}$ and $T_{uA}$ at $O(r^{-2})$. The field redefinitions shift the trace-free part of $h_{AB}$ at $O(r^{-4})$, $V$ at $O(r^{-1})$, the $O(r^{-2})$ part of $g_{uA}$ and $\beta$ at $O(r^{-3})$, and $\tau$ at $O(r^{-3})$; their pure-trace parts and the restoration of the Bondi gauge act at the same or lower orders. Hence $c_{AB}$, $M$, $N_A$, $e_{AB}$ and $\tau_1$ are unchanged, and none of the operators of $S_{\rm other}$ contributes to the trace-free angular equation at the order where $e_{AB}$ evolves, nor to any balance law. These statements hold within the effective action~\eqref{eq:S10} to first order in $\alpha'$.

\section{Leading components at null infinity}
\label{app:components}

We list the leading components used in the text, on the background~\eqref{eq:bondi}--\eqref{eq:matterfalloff}, with $g^{ur}=-1+O(r^{-2})$, $g^{rr}=1+O(r^{-1})$, $g^{AB}=q^{AB}/r^2+O(r^{-3})$ and $\epsilon_{ur\theta\varphi}=r^2\sin\theta+O(r)$:
\begin{gather}
\partial_u\phi=\frac{\dot\phi_1}{r}+O(r^{-2}),\qquad \partial_r\phi=-\frac{\phi_1}{r^2}+O(r^{-3}),\qquad (\nabla\phi)^2=\frac{2\phi_1\dot\phi_1}{r^3}+O(r^{-4}),\\
F_{ur}=\frac{Q}{r^2},\qquad F_{uA}=\dot A^{(0)}_A+O(r^{-1}),\qquad F_{rA}=O(r^{-2}),\qquad F_{AB}=f\epsilon_{AB}+O(r^{-1}),\\
{}^\star F_{ur}=\frac{f}{r^2}+O(r^{-3}),\qquad {}^\star F_{uA}=\epsilon_A{}^B\dot A^{(0)}_B+O(r^{-1}),\qquad
F_{\mu\nu}F^{\mu\nu}=O(r^{-4}),\qquad F\tilde F=O(r^{-4}),\\
F_{u\rho}F_u{}^\rho=\frac{q^{AB}\dot A^{(0)}_A\dot A^{(0)}_B}{r^2}+O(r^{-3}),\qquad
F_{u\rho}F_A{}^\rho=\frac{Q\dot A^{(0)}_A+\dot A^{(0)C}F^{(0)}_{AC}}{r^2}+O(r^{-3}).
\end{gather}
The identities $q^{AB}\psi_A\bar\psi_B=2(\im\tau_0)^2q^{AB}\dot A^{(0)}_A\dot A^{(0)}_B$ and $\re(\bar\Qc\psi_A)=(\im\tau_0)^2\big(Q\dot A^{(0)}_A+\dot A^{(0)C}F^{(0)}_{AC}\big)$ follow from Eq.~\eqref{eq:psi} and $\epsilon_A{}^C\epsilon_C{}^B=-\delta_A^B$, and Eq.~\eqref{eq:curlid} from $\epsilon^{AB}\psi_B=i\psi^A$. Under Eq.~\eqref{eq:SL2}, $\Psi\to\Psi/(c\tau+d)$ follows from $G+\tau F\to(aG-bF)+\frac{a\tau+b}{c\tau+d}(-cG+dF)=\frac{G+\tau F}{c\tau+d}$. The transformation of $\tau_1$ and $\tau_2$ follows from expanding the M\"obius map around $\tau_0$. With Eq.~\eqref{eq:qe},
\begin{equation}
\begin{aligned}
\bar\Qc_1&=-i\,\mathsf q_{e1}-i\,\tau_0f_1,\\
\im\big(\bar\Qc_1\Qc_2\big)&=\im\big[(\mathsf q_{e1}+\tau_0 f_1)(\mathsf q_{e2}+\bar\tau_0 f_2)\big]=\im\tau_0\,\big(f_1\mathsf q_{e2}-\mathsf q_{e1}f_2\big)=(\im\tau_0)^2\big(f_1Q_2-Q_1f_2\big),
\end{aligned}
\end{equation}
and since $(\mathsf q_e,f)$ transforms as a doublet under Eq.~\eqref{eq:SL2}, Eq.~\eqref{eq:DSZ} is invariant. For the low-velocity expansion of Sec.~\ref{sec:spin} we use $K=(1-v^2)\sum_{k\ge0}(k+1)(\hat n\cdot\mathbf v)^k$, whose $\ell=2$ part at second order is $3\,\hat n_{\langle i}\hat n_{j\rangle}v^iv^j$, and the eigenvalue $\tfrac14\ell^2(\ell+1)^2\big[-(\ell-1)(\ell+2)\big]=-36$ at $\ell=2$.

\section{Curvature tensors of the Gauss--Bonnet and Pontryagin couplings}
\label{app:curvature}

\emph{Conventions.} We use $\epsilon_{ur\theta\varphi}=+\sqrt{-g}$, so that $\epsilon^{ur\theta\varphi}=-1/\sqrt{-g}$ and, on the unit sphere, $\epsilon_{\theta\varphi}=\sin\theta$. The dual Riemann tensor is ${}^\star R^{\tau\mu\nu\sigma}=\tfrac12\epsilon^{\nu\sigma\alpha\beta}R^{\tau\mu}{}_{\alpha\beta}$, the Pontryagin density is $\Po=R_{\mu\nu\rho\sigma}\tilde R^{\mu\nu\rho\sigma}$ with $\tilde R^{\mu\nu\rho\sigma}=\tfrac12\epsilon^{\mu\nu\alpha\beta}R_{\alpha\beta}{}^{\rho\sigma}$, and $P_{\mu\alpha\nu\beta}$ is the double dual of Eq.~\eqref{eq:H}.

\emph{Divergence identities.} In four dimensions the double dual is divergence free, $\nabla^\mu P_{\mu\alpha\nu\beta}=0$, by the contracted Bianchi identity. Hence $\nabla^\mu H_{\mu\nu}=P_{\mu\alpha\nu\beta}\nabla^\mu\nabla^\alpha\nabla^\beta f=\tfrac12P_{\mu\alpha\nu\beta}R^{\mu\alpha\beta}{}_\lambda\nabla^\lambda f$, and the four-dimensional identity $P_{\mu\alpha\nu\beta}R^{\mu\alpha\beta}{}_\lambda=-\tfrac14\GB\,g_{\nu\lambda}$, the trace of which is $P_{\mu\alpha\nu\beta}R^{\mu\alpha\nu\beta}=\GB$~\cite{Lanczos:1938sf}, gives the first of Eqs.~\eqref{eq:HCidentities}. The second is the standard divergence identity of the Pontryagin coupling~\cite{Jackiw:2003pm,Alexander:2009tp}, written in the present orientation. We verified both identities by direct computation: the first on a static spherically symmetric metric with arbitrary functions $g_{tt}(r)$, $g_{rr}(r)$ and arbitrary $f(r)$, and the second on a slowly rotating metric to first order in the rotation, with an arbitrary $\vartheta(r,\theta)$, which is the simplest configuration with $\Po\neq0$.

\emph{Coefficients.} For an action $\frac1{16\pi}\int\sqrt{-g}\,[R+\mathcal L_m+\lambda_{\GB}f\GB+\lambda_{\Po}\vartheta\Po]$, diffeomorphism invariance implies that the divergence of the metric equation vanishes when the scalar equations hold. The scalar equations contain $\lambda_{\GB}f'\GB$ and $\lambda_{\Po}\Po$, and with Eqs.~\eqref{eq:HCidentities} this fixes the metric equation to $G_{\mu\nu}=8\pi T_{\mu\nu}-4\lambda_{\GB}H_{\mu\nu}[f]+4\lambda_{\Po}C_{\mu\nu}[\vartheta]$. For Eq.~\eqref{eq:Salpha}, $\lambda_{\GB}=\lambda_{\Po}=\alpha'/8$, $f=e^{-2\phi}$ and $\vartheta=a$, which gives Eq.~\eqref{eq:einstein}. For instance, the dilaton part of $8\pi\nabla^\mu T_{\mu\nu}$ is $2\nabla_\nu\phi\,\Box\phi\supset\tfrac{\alpha'}{8}e^{-2\phi}\GB\nabla_\nu\phi$, which cancels $-\tfrac{\alpha'}{2}\nabla^\mu H_{\mu\nu}=\tfrac{\alpha'}{16}\GB\nabla_\nu e^{-2\phi}$; the axion part of $8\pi\nabla^\mu T_{\mu\nu}$ contains $-\tfrac{\alpha'}{16}\Po\nabla_\nu a$, which cancels $\tfrac{\alpha'}{2}\nabla_\mu C^\mu{}_\nu=\tfrac{\alpha'}{16}\Po\nabla_\nu a$.

\emph{Order-by-order structure.} On the asymptotic solution, the coordinate components of the scalar Hessian are $\nabla_r\nabla_rf=O(r^{-3})$, $\nabla_u\nabla_rf=O(r^{-2})$, $\nabla_u\nabla_uf=O(r^{-1})$, $\nabla_A\nabla_rf=O(r^{-2})$, and with the inverse metric the dominant contravariant component is $f^{uu}=(g^{ur})^2\nabla_r\nabla_rf$. Evaluating each contraction in $H^{\rm TF}_{AB}=P_{AcBd}f^{cd}$ on the asymptotic solution with both polarizations of the shear, only $P_{AuBu}f^{uu}$ reaches $O(r^{-2})$; the contractions with $f^{ur}$, $f^{rr}$, $f^{uC}$ and $f^{CD}$ start at $O(r^{-3})$ or below. The same pattern holds for $C^{\rm TF}_{AB}$, whose leading term comes from $\nabla_\sigma\nabla_\tau\vartheta\,{}^\star R^{\tau(\mu\nu)\sigma}$ with $\sigma=\tau=u$ upstairs, while the term $\nabla_\sigma\vartheta\,\epsilon^{\sigma\alpha\beta(\mu}\nabla_\beta R^{\nu)}{}_\alpha$ contains the derivative of the Ricci tensor, which is $O(r^{-2})$ in its largest component, and is subleading.

\bibliography{heterotic_memory}
\bibliographystyle{elsarticle-num}
\end{document}